%% file: qaSoLidpaper.tex
\documentclass[a4paper,11pt]{article}
\pdfoutput=1 

\usepackage{jinstpub} 
\usepackage[normalem]{ ulem }
\usepackage{soul}
\usepackage{xspace}
\usepackage{textcomp}
\usepackage{amsmath}
\newcommand{\GEANTfour} {{\textsc{Geant4}}\xspace}
\usepackage{lineno}

\title{\boldmath Development of a Quality Assurance Process for the SoLid Experiment}

\input{sections/SoLid_author_list_feb2018.tex}

\abstract{ The SoLid experiment has been designed to search for an oscillation pattern induced by a light sterile neutrino state, utilising the BR2 reactor of SCK$\bullet$CEN, in Belgium. 

 The detector leverages a new hybrid technology, utilising two distinct scintillators in a cubic array, creating a highly segmented detector volume.
 A combination of 5 cm cubic polyvinyltoluene cells, with $^6$LiF:ZnS(Ag) sheets on two faces of each cube, facilitate reconstruction of the neutrino signals. 
Whilst the high granularity provides a powerful toolset to discriminate backgrounds; by itself 
the segmentation also represents a challenge in terms of homogeneity and calibration, for a consistent detector response.
The search for this light sterile neutrino implies a sensitivity to distortions of around $\mathcal{O}$(10)\% in the energy spectrum of reactor $\overline{\nu}_e$. Hence, a very good neutron detection efficiency, light yield and homogeneous detector response are critical for data validation. The minimal requirements for the SoLid physics program are a light yield  and a neutron detection efficiency larger than 40 PA/MeV/cube and 50 \% respectively. 
In order to guarantee these minimal requirements, the collaboration developed a rigorous quality assurance process for all 12800 cubic cells of the detector. 
To carry out the quality assurance process, an automated calibration system 
called CALIPSO was designed and constructed. CALIPSO provides precise, automatic placement of 
radioactive sources in front of each cube of a given detector plane (16$\times$16 cubes). A combination of $^{22}$Na, $^{252}$Cf and  AmBe  gamma and neutron sources
were used by CALIPSO during the quality assurance process. 
Initially, the scanning identified defective components allowing for repair during initial construction of the SoLid detector. Secondly, 
a full analysis of the calibration data revealed initial estimations for the light yield of over 60 PA/MeV and neutron reconstruction efficiency of 68\%, 
validating the SoLid physics requirements.}


\begin{document} 

\maketitle

\flushbottom

\input{sections/Introduction.tex}
\input{sections/Cubes.tex}

\input{sections/Calipso.tex}

\input{sections/LY.tex}
\input{sections/Neff.tex}

\input{sections/Conclusions.tex}

\appendix

\acknowledgments

This work was supported by the following funding agencies: Agence Nationale de la Recherche
grant ANR-16CE31001803, Institut Carnot Mines, CNRS/IN2P3 et Region Pays de Loire, France;
FWO-Vlaanderen and the Vlaamse Herculesstichting, Belgium; The U.K. groups acknowledge
the support of the Science \& Technology Facilities Council (STFC), United Kingdom; We are
grateful for the early support given by the sub-department of Particle Physics at Oxford and High
Energy Physics at Imperial College London. We thank also our colleagues, the administrative
and technical staffs of the SCK$\bullet$CEN for their invaluable support for this project. Individuals have
received support from the FWO-Vlaanderen and the Belgian Federal Science Policy Office (BelSpo)
under the IUAP network programme; The STFC Rutherford Fellowship program and the European
Research Council under the European Union's Horizon 2020 Programme (H2020-CoG)/ERC Grant
Agreement n. 682474 (A. Vacheret); Merton College Oxford.


\bibliographystyle{JHEP}
\bibliography{biblio1}

\end{document}

%% file: sections/SoLid_author_list_feb2018.tex
\collaboration{The SoLid Collaboration}
\author[a]{Y.~Abreu,}
\author[i]{Y.~Amhis,}
\author[d]{G.~Ban,}
\author[a]{W.~Beaumont,}
\author[o]{S.~Binet,}
\author[i]{M.~Bongrand,}
\author[i]{D.~Boursette,}
\author[j]{B.~C.~Castle,}
\author[o]{H.~Chanal,}
\author[b]{K.~Clark,}
\author[k]{B.~Coup\'e,}
\author[o]{P.~Crochet,}
\author[b]{D.~Cussans,}
\author[a,e]{A.~De Roeck,}
\author[d]{D.~Durand,}
\author[h]{M.~Fallot,}
\author[k]{L.~Ghys,}
\author[h]{L.~Giot,}
\author[g]{K.~Graves,}
\author[d]{B.~Guillon,}
\author[h]{D.~Henaff,}
\author[g]{B.~Hosseini,}
\author[g]{S.~Ihantola,}
\author[i]{S.~Jenzer,}
\author[k]{S.~Kalcheva,}
\author[c]{L.~N.~Kalousis,}
\author[f]{M.~Labare,}
\author[d]{G.~Lehaut,}
\author[b]{S.~Manley,}
\author[i]{L.~Manzanillas,}
\author[k]{J.~Mermans,}
\author[f]{I.~Michiels,}
\author[o]{S.~Monteil,}
\author[f,k]{C.~Moortgat,}
\author[b,m]{D.~Newbold,}
\author[l]{J.~Park,}
\author[d]{V.~Pestel,}
\author[b]{K.~Petridis,}
\author[a]{I.~Pi\~nera,}
\author[k]{L.~Popescu,}
\author[f]{D.~Ryckbosch,}
\author[j]{N.~Ryder,}
\author[g]{D.~Saunders,}
\author[i]{M.-H.~Schune,}
\author[h]{M.~Settimo,}
\author[i,n]{L.~Simard,}
\author[g]{A.~Vacheret,}
\author[f]{G.~Vandierendonck,}
\author[k]{S.~Van Dyck,}
\author[c]{P.~Van Mulders,}
\author[a]{N.~van Remortel,}
\author[a,c]{S.~Vercaemer,}
\author[a]{M.~Verstraeten,}
\author[h]{B.~Viaud,}
\author[j,m]{A.~Weber,}
\author[h]{F.~Yermia}

\affiliation[a]{Universiteit Antwerpen, Antwerpen, Belgium}
\affiliation[b]{University of Bristol, Bristol, UK}
\affiliation[c]{Vrije Universiteit Brussel, Brussel, Belgium}
\affiliation[d]{Normandie Univ, ENSICAEN, UNICAEN, CNRS/IN2P3, LPC Caen, 14000 Caen, France}
\affiliation[e]{CERN, 1211 Geneva 23, Switzerland}
\affiliation[f]{Universiteit Gent, Gent, Belgium}
\affiliation[g]{Imperial College London, Department of Physics, London, United Kingdom}
\affiliation[h]{Universit\'e de Nantes, IMT Atlantique, CNRS, Subatech, France}
\affiliation[i]{LAL, Univ Paris-Sud, CNRS/IN2P3, Universit\'e Paris-Saclay, Orsay, France}
\affiliation[j]{University of Oxford, Oxford, UK}
\affiliation[k]{SCK-CEN, Belgian Nuclear Research Centre, Mol, Belgium}
\affiliation[l]{Center for Neutrino Physics, Virginia Tech, Blacksburg, Virginia, 24061, USA}
\affiliation[m]{STFC, Rutherford Appleton Laboratory, Harwell Oxford, United Kingdom}
\affiliation[n]{Institut Universitaire de France, F-75005 Paris, France}
\affiliation[o]{Universit\'e Clermont Auvergne, CNRS/IN2P3, LPC, F-63000 Clermont-Ferrand, France.}
\emailAdd{manzanillas@lal.in2p3.fr, pestel@lpccaen.in2p3.fr}

%% file: sections/Introduction.tex
\section{Introduction}
\label{sec:intro}
Previous short baseline (anti)-neutrino oscillation experiments have evidenced deficits in the observed number of (anti)-neutrino events,
with respect to theoretical expectations \cite{Aguilar:2001ty,Giunti:2010zu,Mention:2011rk}. 

The most relevant discrepancy for SoLid is the Reactor Antineutrino Anomaly (RAA) \cite{Mention:2011rk}, which was first determined from re-evaluation of the antineutrino 
flux and spectra of nuclear reactors  \cite{Mueller:2011nm,Huber:2011wv}; and then confirmed by the 
RENO, Double Chooz, and Daya Bay experiments \cite{An:2016srz}. A possible interpretation of these deficits is the existence of oscillation effects induced by a light sterile neutrino state \cite{Gariazzo:2017fdh}. 
In recent years many experiments have been proposed and constructed in order to test the 
light sterile neutrino hypothesis as the origin of the RAA \cite{Ko:2016owz,Alekseev:2016llm,Serebrov:2017bwe,Almazan:2018wln,Abreu:2018pxg,Ashenfelter:2018zdm}, and have also revealed a significant distortion in the $\overline{\nu}_{e}$ energy spectrum around 5 MeV \cite{Abe:2015rcp,RENO:2015ksa,An:2015nua,Huber:2016xis}. First hints suggest that this distortion is correlated with the reactor power \cite{An:2015nua} and might be associated with the $^{235}$U fuel \cite{An:2017osx}. 

Recent analyses of the RAA suggest a sterile neutrino 
with a $\Delta m^{2}_{41}$ around either 1.3 or 1.7 eV$^{2}$ \cite{Dentler:2018sju,Gariazzo:2017fdh}.
The oscillation of $\overline{\nu}_e$'s into this new sterile state would induce distortions in the 
$\overline{\nu}_e$ spectrum measured at very short baselines ($L<10$ m), which should not 
exceed the 10\% level \cite{Dentler:2018sju}.

The SoLid collaboration have developed a novel detection technique, utilising a highly granular, hybrid detector. As in all oscillation based reactor neutrino experiments, electron antineutrinos are detected by SoLid through the Inverse Beta Decay (IBD) process:
\begin{equation}
 \overline{\nu}_e + p \rightarrow e^{+} +n.
\end{equation}
\label{IBD_eq}
Within the detector volume of SoLid, polyvinyltoluene (PVT) cubes of 5$\times$5$\times$5 cm$^3$ are combined with   
$^6$LiF:ZnS(Ag) screens, on two faces of each cube 
as shown in figure \ref{fig:cubes}, to form volumetric pixels, or ``voxels''.

The PVT serves not only as the target for the IBD reaction, but also as a calorimeter for the positron, which produces a ``prompt'' signal; and as a moderator for the neutron of the IBD reaction. After thermalisation, the neutron will be captured and detected 
in the second scintillator, the $^6$LiF:ZnS(Ag) screens, taking advantage of its high 
neutron capture cross section, initiating a ``delayed'' signal.

After neutron capture, the Li nuclei will break-up, according to the following reaction:
\begin{equation}
 ^{6}_{3}\text{Li} + n \rightarrow \text{ } ^{3}_{1}\text{H} + \alpha
\end{equation}
\label{Li_eq}
which has a Q-value of 4.78 MeV, which will predominantly be deposited in the surrounding ZnS scintillator.

Since both scintillators have distinct light emission properties, a powerful Pulse Shape Discrimination (PSD) algorithm can be used during analyses of events.
The PSD is used in conjunction with a time coincidence between the prompt and delayed signals, to reconstruct the IBD candidate events.  

The $\overline{\nu}_e$ energy is determined through 
 reconstruction of the positron energy using the scintillation from the PVT as the prompt, electronic signal (ES); the $^6$LiF:ZnS provides data regarding the nuclear signal (NS).

In order to guarantee optical isolation and to enhance light collection, the cubic voxels are individually wrapped in Tyvek. 
Voxels are arranged into planes of 16$\times$16 cubes, with 10 planes being grouped into one SoLid module.
The SoLid detector includes 5 modules, for a total target mass of 1600 kg. 
Both prompt and delayed signals are read out by the same network of wavelength shifting fibres and  Multi-Pixel Photon Counters (MPPCs) \cite{Hamamatsu2016}.
A detailed description of the detector concept and the performance of the constructed SM1 prototype can be found in \cite{Abreu:2017bpe,Abreu:2018pxg}.

The granularity inherent in the hybrid voxel technology allows for clear identification of the neutrino signals, significantly reducing backgrounds.
Constructing such a segmented detector provides a significant 
challenge in terms of homogeneity, light yield and neutron detection efficiency, which are of paramount importance in reactor-based searches for light sterile neutrinos, where oscillation effects are expected to be small. In order to have the sensitivity to these small effects, the SoLid minimal requirements include a light yield larger than 40 PA/MeV/cube in order to guarantee an energy resolution better than 16\%, and a neutron detection efficiency larger than 50\% to achieve an IBD detection efficiency of about 30\%.

To validate the necessary performance of the SoLid detector and to enable prompt identification of defective components, a quality assurance (QA) process was developed during construction.
To this end, an automated calibration system called CALIPSO was constructed. CALIPSO provided  an initial calibration 
for all 50 planes of SoLid before they were assembled into modules for the BR2 reactor site. 
Initial estimations of the light yield and neutron detection efficiency for all 12800 voxels were obtained and validated using gamma and neutron radioactive sources.

Section \ref{sec:cubes} of this paper describes the frame construction procedure, and section \ref{sec:calipso} describes in detail the design and performance 
of CALIPSO.  
Sections \ref{sec:LY} and \ref{sec:Neutron} detail the methods used to assess the 
light yield and neutron detection efficiency. Finally section \ref{sec:concl} presents 
a discussion and conclusions on the quality assurance process of the SoLid experiment.

%% file: sections/Cubes.tex
\section[Preparation of the SoLid Planes]{Preparation of the SoLid Planes }
\label{sec:cubes}
\begin{figure}[h]
\centering
\begin{tabular}{c c}
\includegraphics[width=0.55\linewidth]{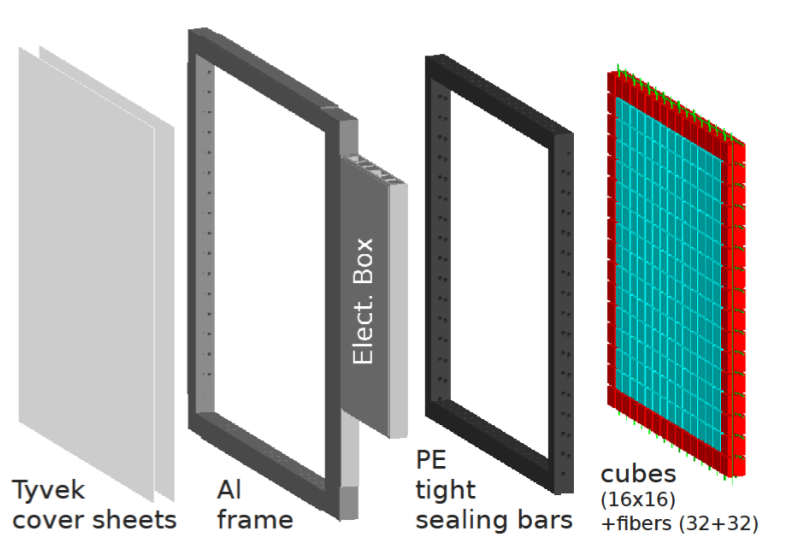} & \includegraphics[width=0.35\linewidth]{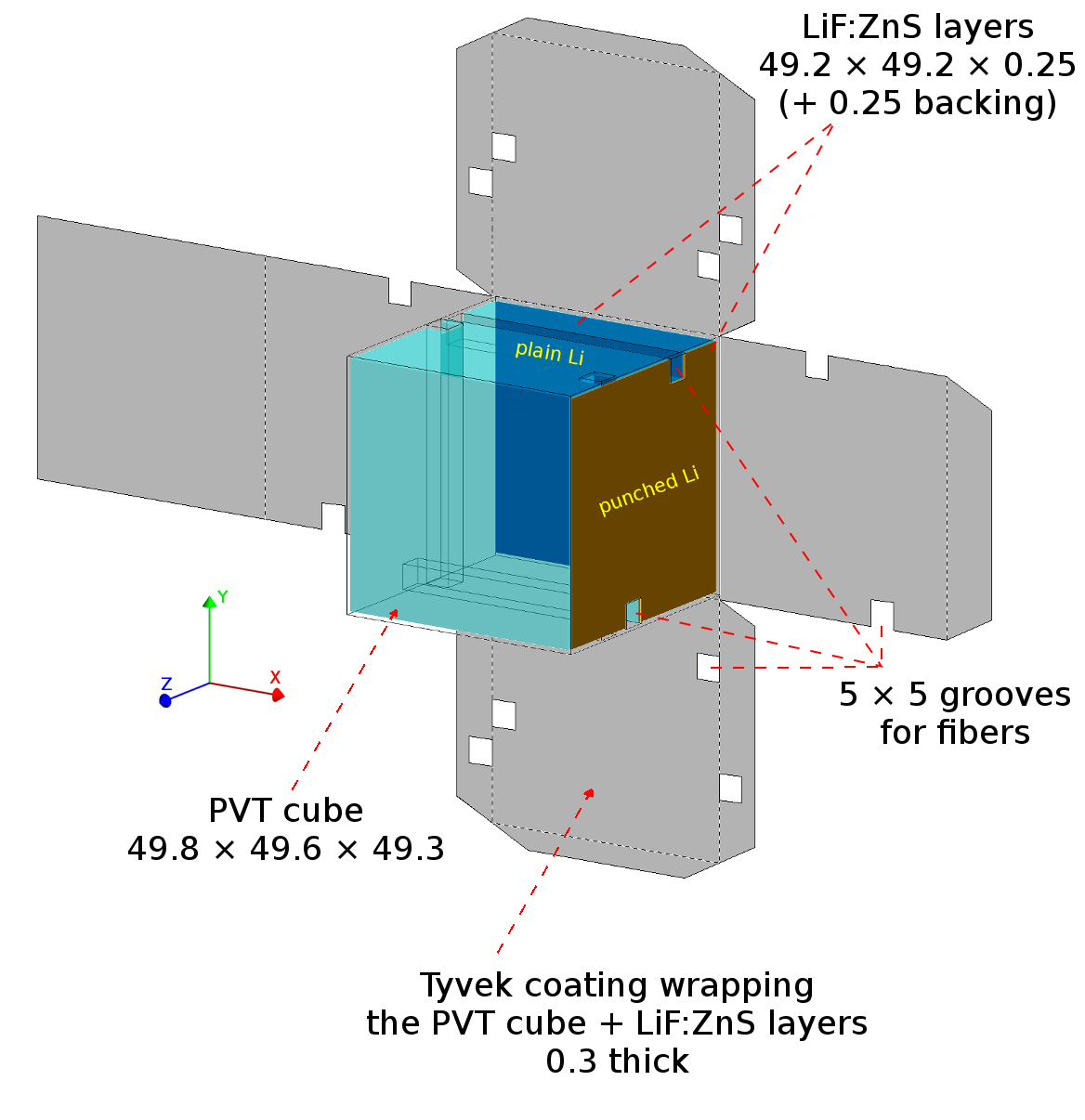}
\end{tabular}
\caption{Left : Different components of a SoLid plane.
Right : Sketch of a fundamental SoLid unit, a voxel.  }
\label{fig:cubes}
\end{figure}

The SoLid detector is composed of 12800 fundamental cubic voxel units.
Each voxel comprises a PVT cube of 5$\times$5$\times$5 cm$^3$ in conjunction with two layers of $\sim$ 250 $\mu$m thick 
$^6$LiF:ZnS(Ag), as illustrated in figure \ref{fig:cubes}. This voxel 
configuration was defined after a dedicated study to improve the light collection of the planar sub-assemblies \cite{Abreu:2018ajc}.

During construction of the voxel units, each PVT cube is cleaned, measured and weighted. Metrics for the lithium sheets and Tyvek wrappings are added to the cube data to form a detailed assay of each voxel. The voxels are then tagged and stored in hermetic plastic boxes until mounting within their specific planes.

For each plane, an aluminium frame provides both mechanical support and a mounting point for the associated electronics. Within this frame a border of 5 cm thick polyethylene (PE) is used as both an internal neutron reflector and as external shielding. A Tyvek sheet is applied to one face of the plane, to offer lateral mechanical support to the voxels, and to provide additional optical isolation between planes. The voxels locations are recorded and linked to the assay database, before 64 optical fibres are threaded throughout the entire plane. This results in four fibres passing through each of the 256 voxels. Each fibre is coupled via 3D printed mountings; at one end to a mirror, and at the other to an MPPC, using optical grease for maximal light transmission. The supports are designed to adhere to the aluminium frames and are glued in place. Prior to cabling the 64 MPPCs to the readout electronics, a final sheet of Tyvek is layered over the open side to fully enclose the voxel array, as in figure \ref{fig:cubes}. 

Following assembly, each completed plane is calibrated in a light-tight room housing the CALIPSO robotic system.

%% file: sections/Calipso.tex
\section[The CALIPSO Calibration System]{The CALIPSO Calibration System}
\label{sec:calipso}
\subsection[CALIPSO Design]{CALIPSO Design}
The CALIPSO system was designed to perform a time efficient and accurate quality assurance process while constructing the SoLid detector.

\begin{figure}[h]
\centering
\begin{tabular}{c c}
 \includegraphics[width=0.9\linewidth, trim={5cm 0 0 0},clip]{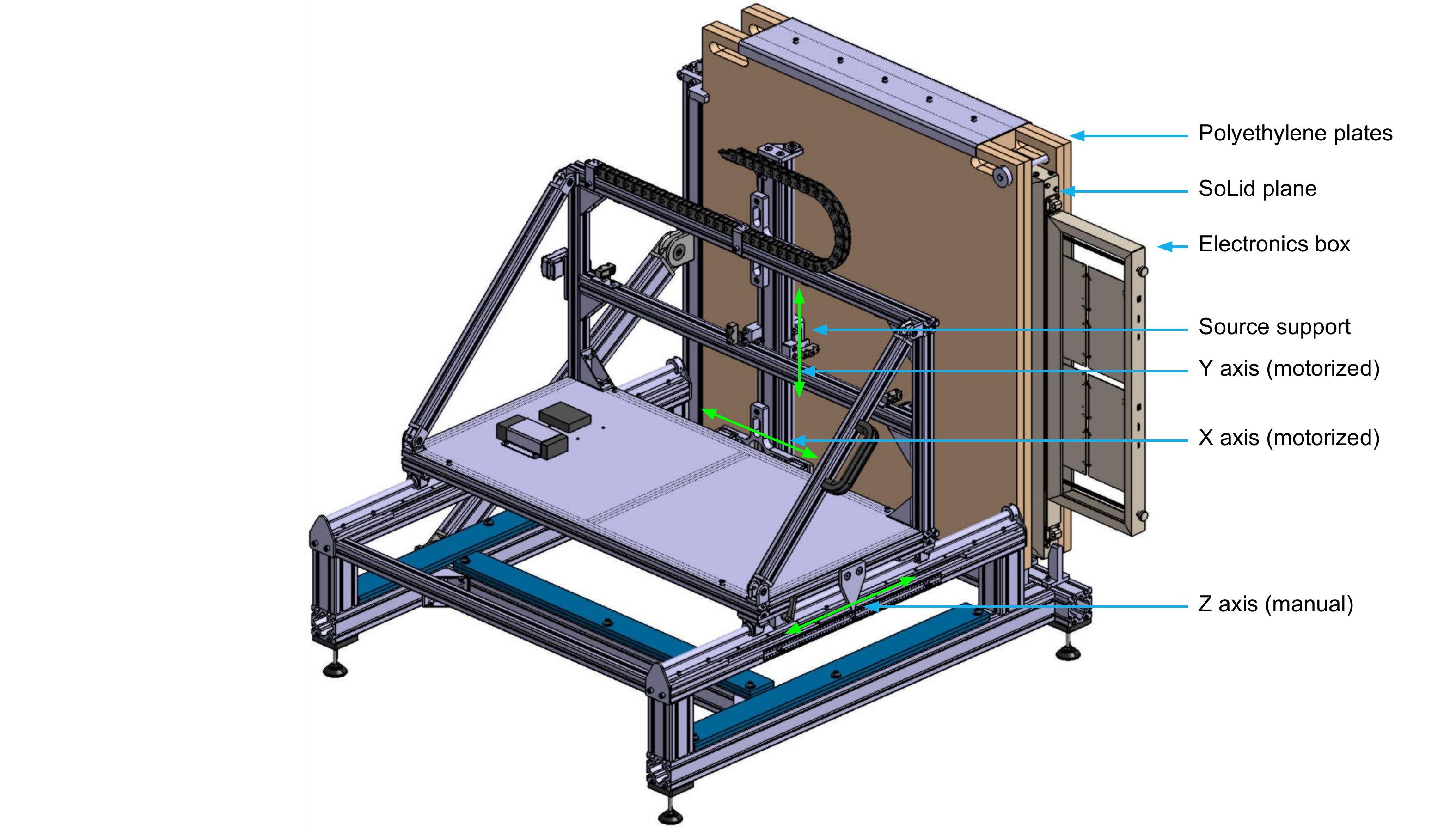}
\end{tabular}
\caption{Illustration of the CALIPSO system for calibration of the SoLid planes, which operates in both neutron and gamma modes.
 The electronic box is placed on the right hand side of the planes.
The CALIPSO robot provides sub-millimetre precision for accurate and consistent placement of the 
radioactive sources at any point in the XY plane.
}
\label{fig:calipso}
\end{figure}

\begin{figure}[!]
\centering
\begin{tabular}{c c c}
  \includegraphics[width=0.27\linewidth]{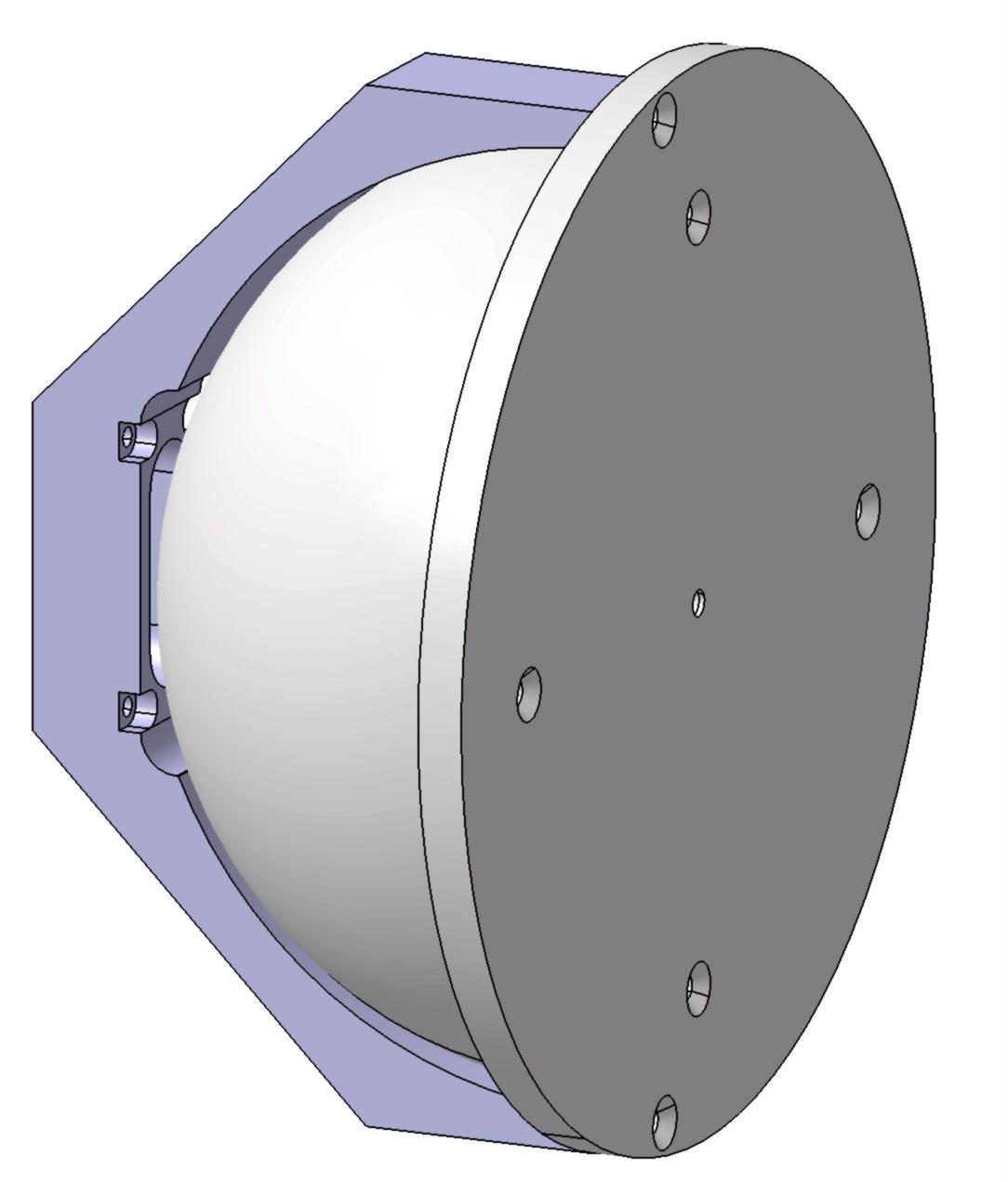} & \includegraphics[width=0.35\linewidth]{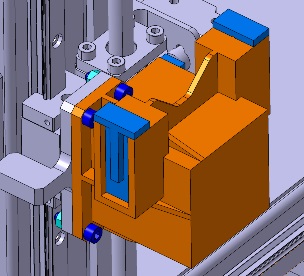} 
\end{tabular}
\caption{Left : Neutron collimator. Right : 3D external head for gamma calibration with $^{22}$Na source.}
\label{fig:calibrationheads}
\end{figure}

The system is driven by a dedicated data acquisition system, which provides simultaneous control of the robots movement and data taking.
CALIPSO has sub-millimetre precision $\mathcal{O}$(0.5) mm in the XY axes with adjustment in the Z axis being provided by a graduated rolling chassis, upon which the robotic armature is mounted. 
The CALIPSO system has been designed to operate in both neutron and gamma modes.
In neutron mode the planes are placed between plates of polyethylene (brown plates in the figure \ref{fig:calipso}) and a neutron source is placed
inside a polyethylene collimator (see figure \ref{fig:calibrationheads}). In gamma mode, these PE plates are removed and a system for an external trigger is used (see figure \ref{fig:calibrationheads}).


\subsection[CALIPSO Readout System]{CALIPSO Readout System}
Acquisition of the MPPC signals was performed using a prototype electronics box 
placed on one side of the frames as shown in figure \ref{fig:calipso}. The electronics box is composed of 2 parts:
an analogue and a digital front-end. For the analogue front end, two boards serving 32 channels each, are used in each 
electronics box. These two boards provide each MPPC with 
the corresponding bias voltage, set according to a value controlled by the digital board.
The analogue boards also amplify and shape the signals before transmitting to the digital board. The digital board includes the trigger, on a 
Field-Programmable Gate Array (FPGA) chip, and the Analogue-Digital Converter chips. 
The 8 Analogue-Digital Converters
sample 8 channels each, at a rate of 40 MHz with a 14-bit  resolution \cite{Ryder:2016rph,Abreu:2018njy}.

The CALIPSO calibration campaign facilitated testing of the prototype electronics box prior to manufacturing.
Using data acquisition software designed for the fully assembled experiment, CALIPSO also served as a trigger and DAQ development platform.

\subsection[CALIPSO Monte Carlo]{CALIPSO Monte Carlo}
A dedicated Monte-Carlo model (\GEANTfour based \cite{Agostinelli:2002hh}), including CALIPSO and its
direct environment was developed, in order to optimise the setup and the QA procedure.
The $^{22}$Na radioactive gamma source was simulated using the radioactive decay class from \GEANTfour.

Accounting the $^{22}$Na source activity, the exposure time per cube was set to 30 seconds, in order to guarantee at least 15000 $\gamma$ interactions in each PVT cube.
The recorded energy spectrum for each cube is then used 
to extract the light yield per cube using two approaches that are described in section \ref{sec:LY}.

Neutron response was studied using standard commercial AmBe and $^{252}$Cf neutron sources.
The activity reported by the manufacturer at the time of fabrication (November of 2016) of the sources are 37 MBq and 37 kBq for 
the AmBe and $^{252}$Cf respectively \footnote{ For the AmBe source the quoted 
activity is the $\alpha$ activity of $^{241}$Am, and in the case of the $^{252}$Cf the quoted activity includes $\alpha$ + fission rate decays.};
however, the neutron yield is much lower as presented in table \ref{tab:neutronFlux}. 
These sources underwent a dedicated calibration process at the National Physical Laboratory (NPL) in the UK, where uncertainties in the neutron flux of 2\% for 
the AmBe and 1\% for the $^{252}$Cf were determined. This neutron activity has been used as reference for the Monte-Carlo estimations. 
An activity correction was applied using the half life of the two isotopes of 432.6 years for AmBe and 2.645 years for $^{252}$Cf.
The mean energies of the neutrons emitted by these sources are 4.2 MeV for the AmBe and  2.1 MeV for the $^{252}$Cf, as shown in table \ref{tab:neutronFlux}.
\begin{table}
\begin{center}
\begin{tabular}{|c|c|c|}
\hline Source & AmBe & $^{252}$Cf\\ \hline Activity (n/s) &
$1794\pm35$ & $3763\pm44$\\ \hline E$_{max}$ & 11 MeV & 15
MeV\\ \hline E$_{mean}$ & 4.2 MeV & 2.1 MeV \\ \hline
\end{tabular}
\caption{Neutron activity of sources used during the QA process as calibrated by the National Physical Laboratory
  (UK) on January of 2017. By December of 2017 when the QA was ending the estimated neutron activities were 1791.5 and 2993.2 n/s
  for the AmBe and $^{252}$Cf respectively.}
\label{tab:neutronFlux}
\end{center}
\end{table}

Fast neutrons emitted from the AmBe or $^{252}$Cf sources have a low probability (< 5\%) of being captured directly in the plane mounted on CALIPSO.
Consequently the probability of seeing a neutron captured after scattering from the CALIPSO apparatus is significantly increased.
These scattered neutrons can bias the neutron response. Therefore, a significant reduction in the frequency of fast neutron scatterings from CALIPSO, into the cubes being calibrated, had to be ensured. 
For this purpose, while in neutron mode, CALIPSO was augmented with two polyethylene plates of 5 cm on each side of the plane, as 
previously illustrated in figure \ref{fig:calipso}. These two moderating volumes induce two effects on the fast neutron population detected within the cubes under investigation. Firstly, by moderating and thermalising the 
neutrons coming directly from the source, they increase the probability of seeing a direct interaction in the plane. 
Secondly, the PE plates reduce the probability of seeing an indirect interaction from a fast neutron initially scattered from the CALIPSO structure.

\begin{figure}[!]
\centering
\begin{tabular}{c c}
  \includegraphics[width=0.40\linewidth]{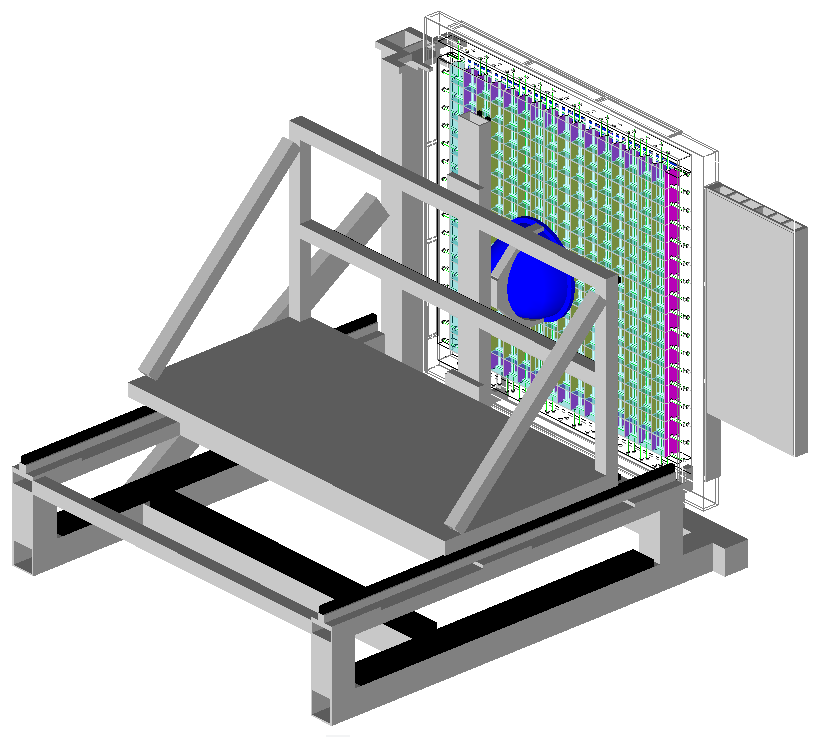} & \includegraphics[width=0.50\linewidth]{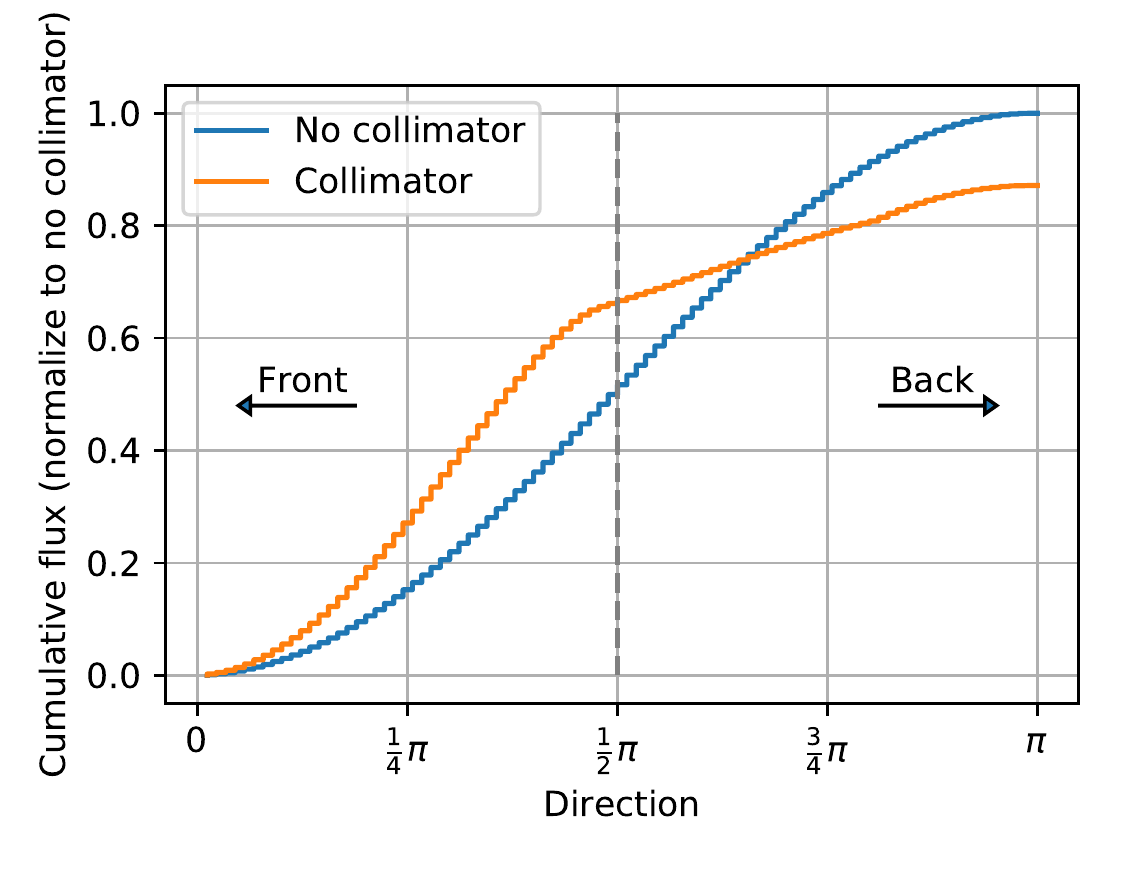}
\end{tabular}
\caption{Left : CALIPSO \GEANTfour Monte Carlo model in neutron mode. The neutron collimator can be observed in blue. Right : effect of the neutron collimator on the 
neutron flux as a function of the angle with respect to the orthogonal axis of the SoLid plane. }
\label{fig:calipsoMC}
\end{figure}

In addition to the fast neutron reduction regime, a neutron collimator was designed to influence the directional neutron flux with respect to the cube under investigation. 

The optimal collimator geometry was determined to increase the neutron detection rate by a factor of 1.4 due to both the forward collimation and from a less
energetic flux (see figure \ref{fig:calipsoMC}). Thus, the number of neutrons entering the detector,
after diffusion in the room, decreased by a factor of 5 to 10,
depending of the position of the source. As an additional step in the analysis, only those cubes close to the source were used for calibration (see section \ref{sec:Neutron}); taking 
into account that the scattered neutrons are more likely detected in cubes further from the source position. 
 
\begin{figure}[!]
\centering
\begin{tabular}{c c}
  \includegraphics[width=0.48\linewidth]{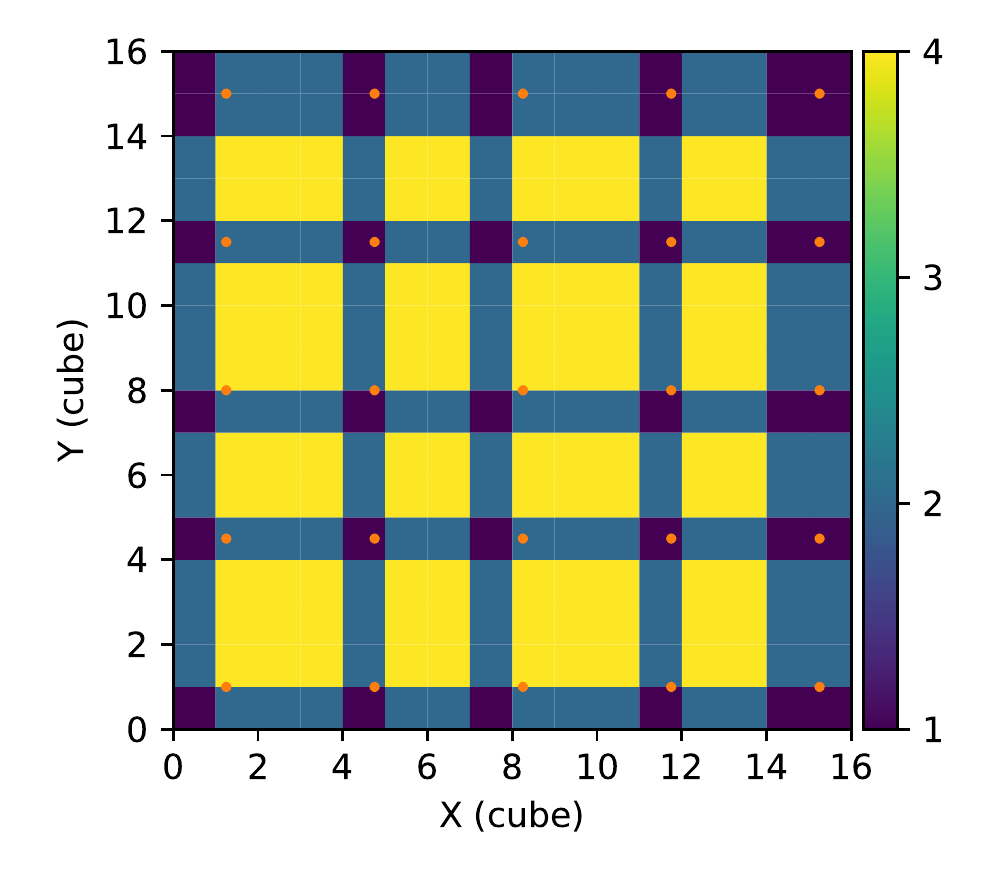} &
  \includegraphics[width=0.48\linewidth]{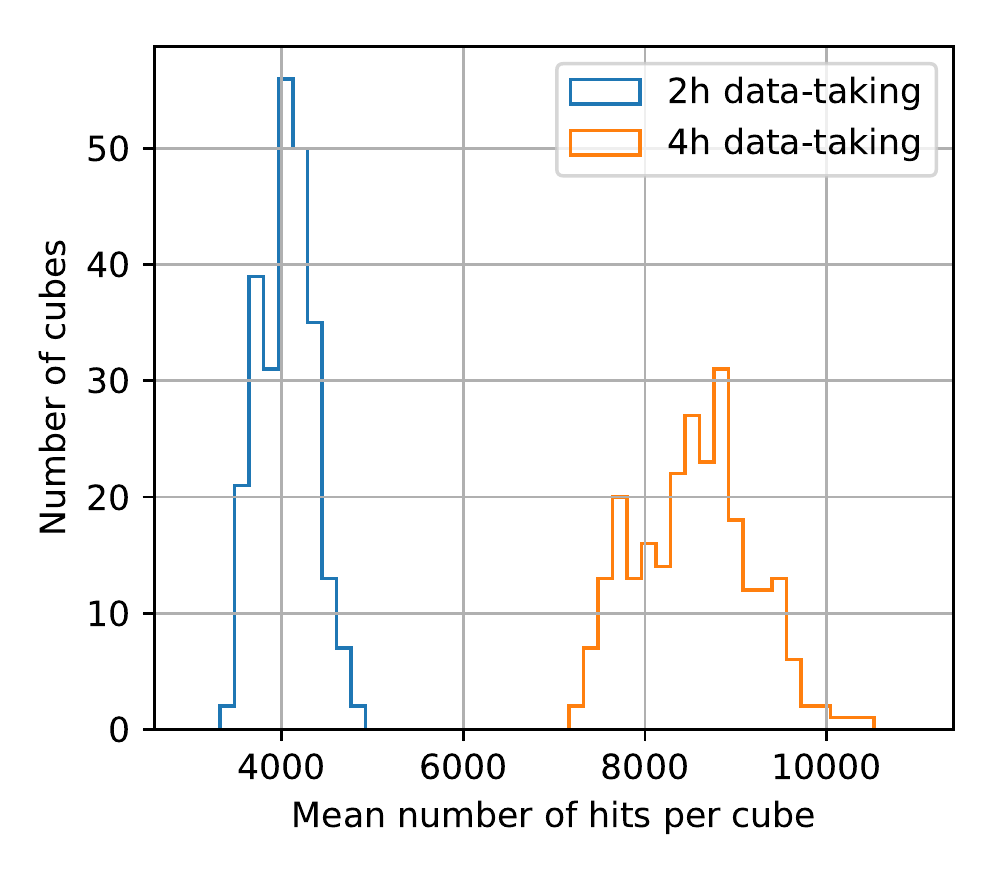}
\end{tabular}
\caption{Left : Number of measurement points contributing to each cube after application of cuts. The neutron 
source is placed in 25 positions as illustrated with the orange dots. Only cubes close to the source are considered to the estimations.   
  Right : Distribution of the mean number of hits per cube for 2 and 4 hours of data taking. A homogeneous distribution is achieved with the 25 
  XY positions in each plane. 
  The average number of hits recorded was 4051.1 ($\sigma = 288.4$) or 8542.9 ($\sigma = 636.5$) per cube for the 2 or 4 hours of data taking.
  }
\label{fig:NS_merge}
\end{figure}

Combining these different actions, a Monte-Carlo study indicated that $\approx$\,98\,\% of the captured neutrons are 
only scattered in the collimator-shielding-plane volume.
In addition, this study showed that placing the source in this configuration for 25 XY positions, as shown in figure \ref{fig:NS_merge},
provides an homogeneous distribution of the NS events among the plane.
This was important for the goal of accumulating at least
4000 neutron captures in each cube.
Finally, given that the QA process was scheduled for six months, effects of mechanical shifting within the apparatus were reduced to a minimum. Consequently, this configuration could be maintained over a number of months,
without the need for research personnel to closely monitor the immediate CALIPSO environment.  
Further details of the calibration process are expanded upon in section \ref{sec:Neutron}.

%% file: sections/LY.tex
\section{Light Yield Measurement}
\label{sec:LY}

\begin{figure}[!]
	\centering
	\includegraphics[width=7cm]{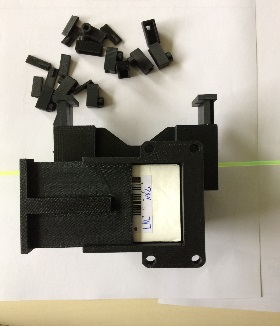}
	\caption{3D printed external head for housing the external trigger assembly and the $^{22}$Na source.}
	\label{fig:backHeadNa}
\end{figure}

In order to assess the light yield (LY) of each SoLid voxel, a $^{22}$Na gamma source was used, in conjunction with an external trigger.
The external trigger consists of a PVT cube, read out by a small wavelength shifting fibre, coupled with an MPPC at each end. The system is contained 
within a 3D printed, externally mounting head; designed to combine the external trigger components, together with the $^{22}$Na source as shown in figure \ref{fig:backHeadNa}.

\begin{figure}[!]
  \centering
  \begin{tabular}{c c}
        \includegraphics[width=0.6 \linewidth]{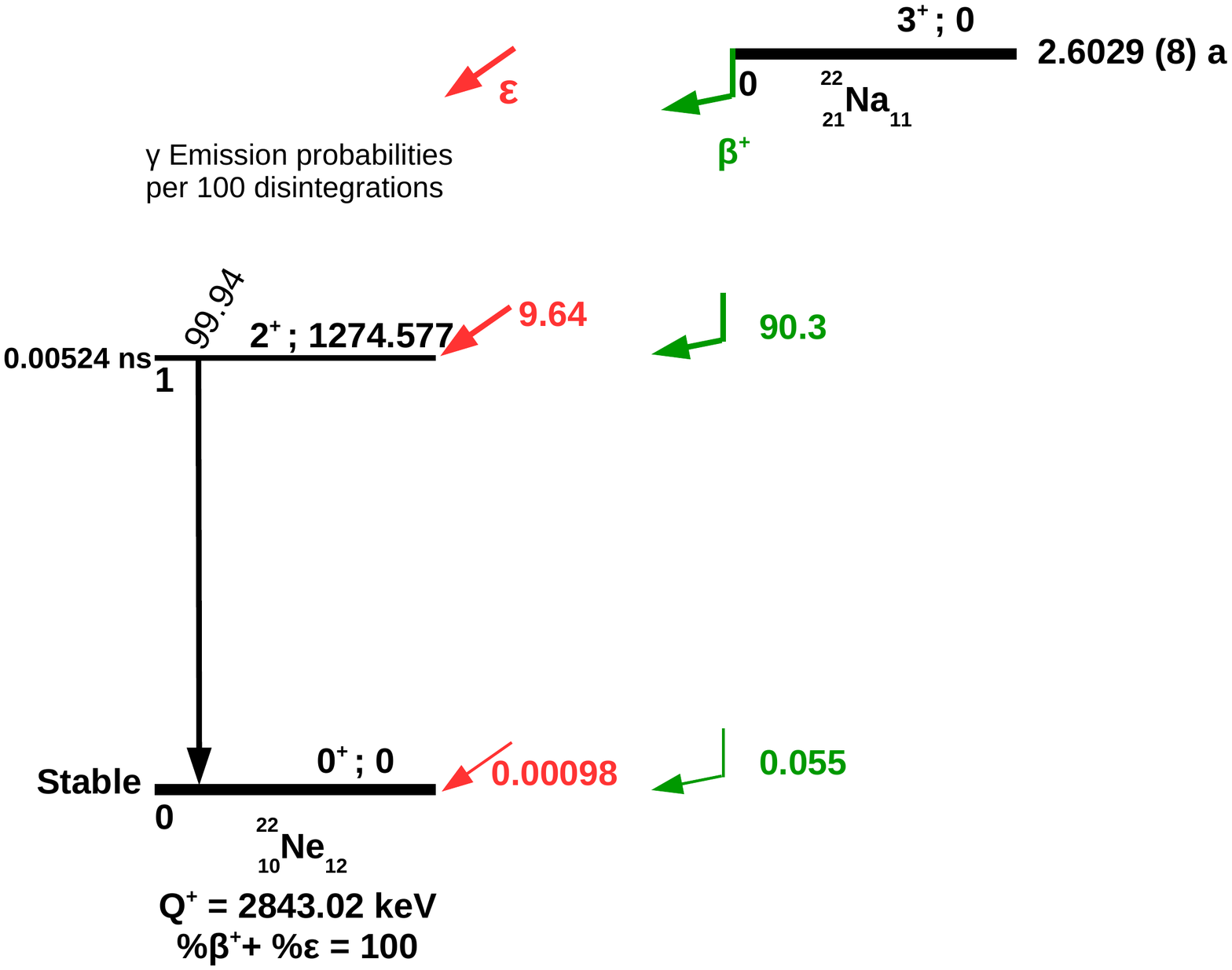}&
        \includegraphics[width=0.3 \linewidth]{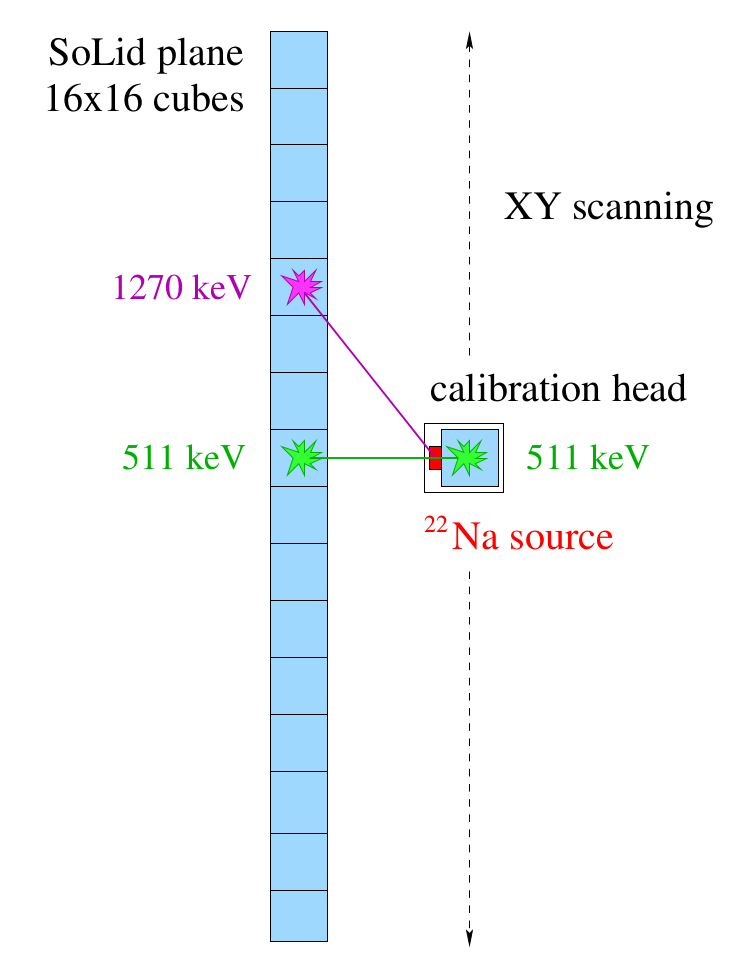}
  \end{tabular}
	\caption{ Left : Decay scheme of a $^{22}$Na source. In about 90\% of cases it decays via $\beta +$ emission to the 
	first excited state of $^{22}$Ne, which in turn goes into its ground state through the emission of a 1.27 MeV gamma.
	Right : Representation of the external trigger system using the 511 keV gammas from the $^{22}$Na source.}
	\label{fig:NaSource}
\end{figure}

$^{22}$Na decays 
via $\beta^{+}$ (90.3\%) and via electronic capture (9.64\%) into $^{22}$Ne as illustrated in figure \ref{fig:NaSource}. In almost all cases,
$^{22}$Na decays into the first excited state of $^{22}$Ne, which in turn decays to its ground state via 
the emission of a 1.27 MeV gamma. Hence, in 90\% of the $^{22}$Na decays the emission of a positron in conjunction 
with a 1.27 MeV gamma occurs; however, the $e^{+}$ annihilates inside the source capsule, emitting two back-to-back 511 keV gammas. 
These gammas are then used for the external trigger as illustrated in figure \ref{fig:NaSource}.
If one of these gammas interacts in the external cube, 
we deem this a triggered event, and the full SoLid plane is read out. In this way, calibration samples  
with almost zero background can be collected, which allows an accurate calibration using the 1.27 MeV gamma, in conjunction with the 511 keV gammas. 
It is important to note that the activity of the source and its uncertainty have no impact on the estimation of the light yield using this method. However, they will 
affect the signal to background ratio  and time of exposure which can be tuned to accumulate the required statistics. 


\subsection{Determining MPPC Operating Voltage}
Before taking calibration data, the correct operation of 
all 64 MPPCs, within each frame, must be verified.
The validation process began with an initial non-equalised over-voltage data run. 
In a few cases the initial run identified unresponsive MPPCs which were replaced before
continuing with the process. 
As the operating breakdown voltage for each MPPC 
is different, and requiring all the MPPCs to operate with an over voltage (OV) of 1.5 V, 
a voltage scan is required to determine the individual breakdown values.
The voltage scan consisted of 25 runs using a fixed high
voltage, close to the manufacturers nominal value in conjunction with a variable low voltage input. 
In these runs the gain of each MPPC was measured, which increased linearly with the OV.
The individual breakdown voltages were identified by fitting the resulting gain values as a function of the varying voltage,
and extrapolating to a gain of zero. 
For an OV of 1.5 V, a gain of about 22 Analogue-to-Digital Conversion units (ADC) per pixel avalanche was determined.
Uncertainties in the estimation of the breakdown voltage, and the variation in gain response with voltage between the MPPCs, 
translated into the operational gains varying from channel to channel by about 3\%. 
This methodology was further refined during calibration runs after installation at BR2 and achieved an equalisation variance at the 1\% level.

\subsection{Cube Light Yield and Signal Reconstruction}
 
\begin{figure}[!]
	\centering
	\includegraphics[width=\linewidth]{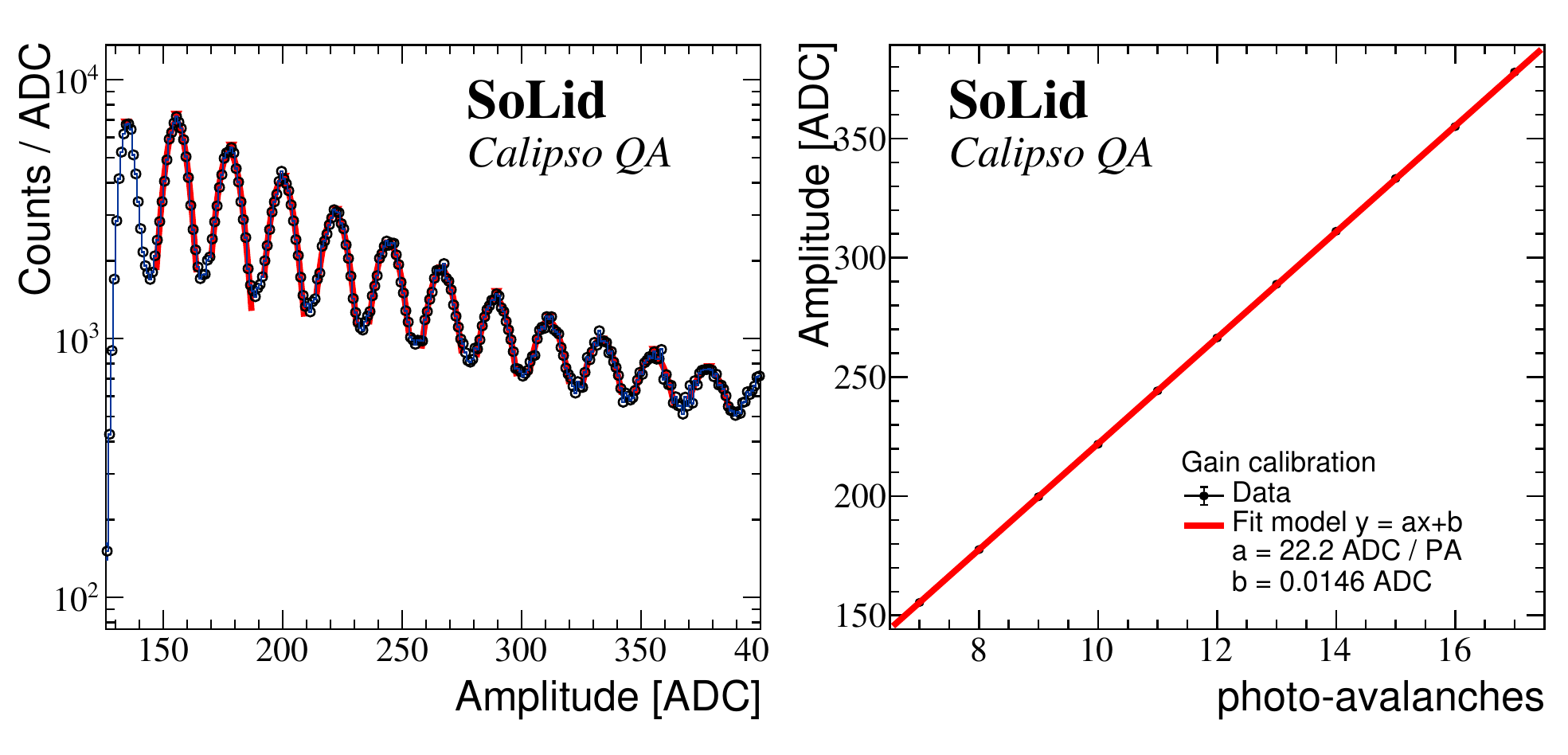}
	\caption{Left : Detected spectrum in an MPPC using the $^{22}$Na source, the PA arrival peaks can be clearly identified.
	The first peak that appears in the spectrum corresponds to the sixth PA, not the first PA. The previous peaks are 
	excluded in the data by setting a zero suppression threshold of 5 PA during the initial QA campaign.\newline
	Right : Linear fit of the PA peaks, where the slope corresponds to the gain. In this particular case the gain, $g$, 
	is calculated at 22.2 $\pm$ 0.01 ADC per PA. The intersection of the curve for 0 PA is 0.01 $\pm$ 0.07 ADC, in very good agreement with a linear response.}
	\label{fig:GainsCalib}
\end{figure}

An additional goal of the quality assurance process was to identify possible defective 
voxels in the SoLid planes. This was carried out  
by calculating the light yield in each voxel from the total amount of light collected in the 4 MPPCs associated with
each cube. Variations in gain from channel
to channel need to be taken into account before summing the signals of the 4 sensors, as each MPPC has a slightly different breakdown response. 
Gains for the individual voxels are recalculated by identifying the PA values in the $^{22}$Na energy spectra; 
each PA distribution was fitted with a Gaussian function as shown in figure \ref{fig:GainsCalib} to obtain the positions of the PA peaks. 
A linear fit of the PA peaks
has a slope corresponding to the gain of each channel.  The parameters of the fit function were tuned 
to operate automatically with the CALIPSO data. 

Figure \ref{fig:GainsCalib} shows the detected amplitude spectrum $A$, in ADCs for 
a channel using the $^{22}$Na source. 
This spectrum is thus used to adjust the gains $g$. Once the gain is determined, it is used to make the conversion from ADC counts
to PA. 
The first peak in figure \ref{fig:GainsCalib} (Left) corresponds to the 6th PA peak, and not the 1 PA associated peak. This 
"shift" is caused by a zero suppression (ZS) threshold set to 5 PA being used during the data taking 
in order to reduce the data rate. This reduction in data rate is needed 
since the QA data taking is performed at standard room temperature
($\sim$25\textdegree), at which the dark rate of peaks below 5 PA is unacceptably high and dominates the data taken.
The SoLid detector will operate at around 10 \textdegree C, with data taking utilising a ZS threshold of 1.5 PA. 
This ZS is driven by the data rate, a ZS  value  at  0.5 PA  will  remove
the  pedestal  contribution,  whilst  retaining  all  SiPM signals, resulting in a waveform compression factor of
around  50.  Increasing  the  threshold  further  to  1.5 PA can  provide  another  order  of  magnitude  of  waveform
compression, at the expense of removing the single PA signals \cite{Abreu:2018njy}.

To reconstruct the total amount of light produced in a given cube, the total $^{22}$Na spectrum per cube must be computed. To effect this, coincidences are sought
between the two vertical and the two horizontal sensors, coupled to the 4 fibres going through each cube.
Thus, the total amplitude per cube $A_{ij}$ in PA is defined as:
\begin{equation}
A_{ij} =  \frac{A^{t}_{i}}{g^{t}_{i}}+ \frac{A^{b}_{i}}{g^{b}_{i}}+ \frac{A^{l}_{j}}{g^{l}_{j}}+\frac{A^{r}_{j}}{g^{r}_{j}} 
\end{equation}
with $t,b,l,r$ for the position of the sensors on the top, bottom, left and right sides of the plane respectively, and $i,j$
for the cube co-ordinates in the plane. 

Gammas from the $^{22}$Na source (511 keV and 1270 keV) interact in the PVT mostly through Compton scattering.
In addition, given the granularity of the detector planes, only a fraction of the total gamma energy 
is deposited within each PVT cube.
Consequently no narrow photo-peak can be reconstructed within individual cubes.
The light yield must therefore be derived from a more complicated distribution, and two approaches were employed to this end. The first method consists of 
fitting the Compton edge profile of the spectrum by an analytical function and  the result compared to the predicted value.
The second method compares the measured energy spectrum to a {\GEANTfour} simulated sample varying the 
light yield and energy resolution. In the next two sections these methods are discussed in more detail.

\subsection{Compton Edge Analytical Fit}

For a given cube undergoing calibration, we assume that a 1.27 MeV gamma interacts within the cube only via Compton scattering and, that the gamma only scatters once per cube. In this instance, the distribution of the true energy deposits of scattered electrons is defined according to the Klein-Nishina cross-section, $\sigma_{c}$  \cite{SICILIANO2008232,KNXSection}:

\begin{equation}
\dfrac{d\sigma_{c}}{dT}=\dfrac{\pi r^{2}_{e}}{m_{e}c^{2}\alpha^{2}} \left( 2+ \left( \dfrac{T}{E_{0}-T} \right)^{2} \left( \frac{1}{\alpha^{2}}+\frac{E_{0}-T}{E_{0}} -\frac{2}{\alpha}\left( \dfrac{E_{0}-T}{T} \right) \right) \right)
\label{eqKN}
\end{equation}
where $T$ represents the kinetic energy of scattered electrons, $\alpha=E_{0}/m_ec^{2}$, $m_{e}$ the electron mass and $E_{0}$ the initial energy of the incident photon.
This cross section is peaked for energies approaching the kinematical  limit for the energy transferred to the scattered electron, and displays an abrupt
fall to zero above this energy. This \textit{Compton Edge} (CE) is the strongest feature of this distribution and  is used to determine the light yield.
Determining the CE position in the distribution of PAs, can be translated into a light yield since the theoretical CE is well known.
The theoretical CE is computed using an angle of deflection, $\theta$, of the incident gamma of 180 degrees.
This leads to the following equation:
\begin{equation}
E_{CE}=E_{T}(max)=E_{0}  \left( 1-\dfrac{1}{1+\frac{2E_{0}}{m_{e}c^{2}}} \right) 
\label{ComptonEdgeEq}
\end{equation}
wich only depends of the initial energy of the incident gamma $E_{0}$ and the electron mass $m_{e}$. 
For gammas of 511 keV and 1270 keV the Compton edges are 341 keV and 1057 keV
respectively. 

Experimentally the Compton edge is smeared according to the energy resolution of the detector. 
The CE value can be identified by using the shape of the energy spectrum; then fitting this spectrum profile according to equation \ref{eqKN}, convoluted with a Gaussian function, in order to account for a stochastic energy resolution.

The resulting function can be written as:
\begin{equation}
f(x) = \int_{0}^{E_{CE}}\dfrac{d\sigma_{c}}{dT} (T) \dfrac{1}{\sqrt{2\pi}\sigma_{0}\sqrt{T}}e^{-0.5\dfrac{(x-T)^{2}}{\sigma^{2}_{0}T}} dT
\label{eqKNConvoluted}
\end{equation}
And the necessary normalisation term to obtain a probability density function (pdf) is given by:
\begin{equation}
\int^{\infty}_{-\infty}f(x)dx = \int_{0}^{E_{CE}}\dfrac{d\sigma_{c}}{dT} dT
\label{eqKNpdf}
\end{equation}
where $E_{CE}$ and $\sigma_{0}$ are respectively the Compton edge and the energy resolution.

\begin{figure}[!]
	\centering
	\includegraphics[trim={0 1cm 0 0},width=0.7 \linewidth]{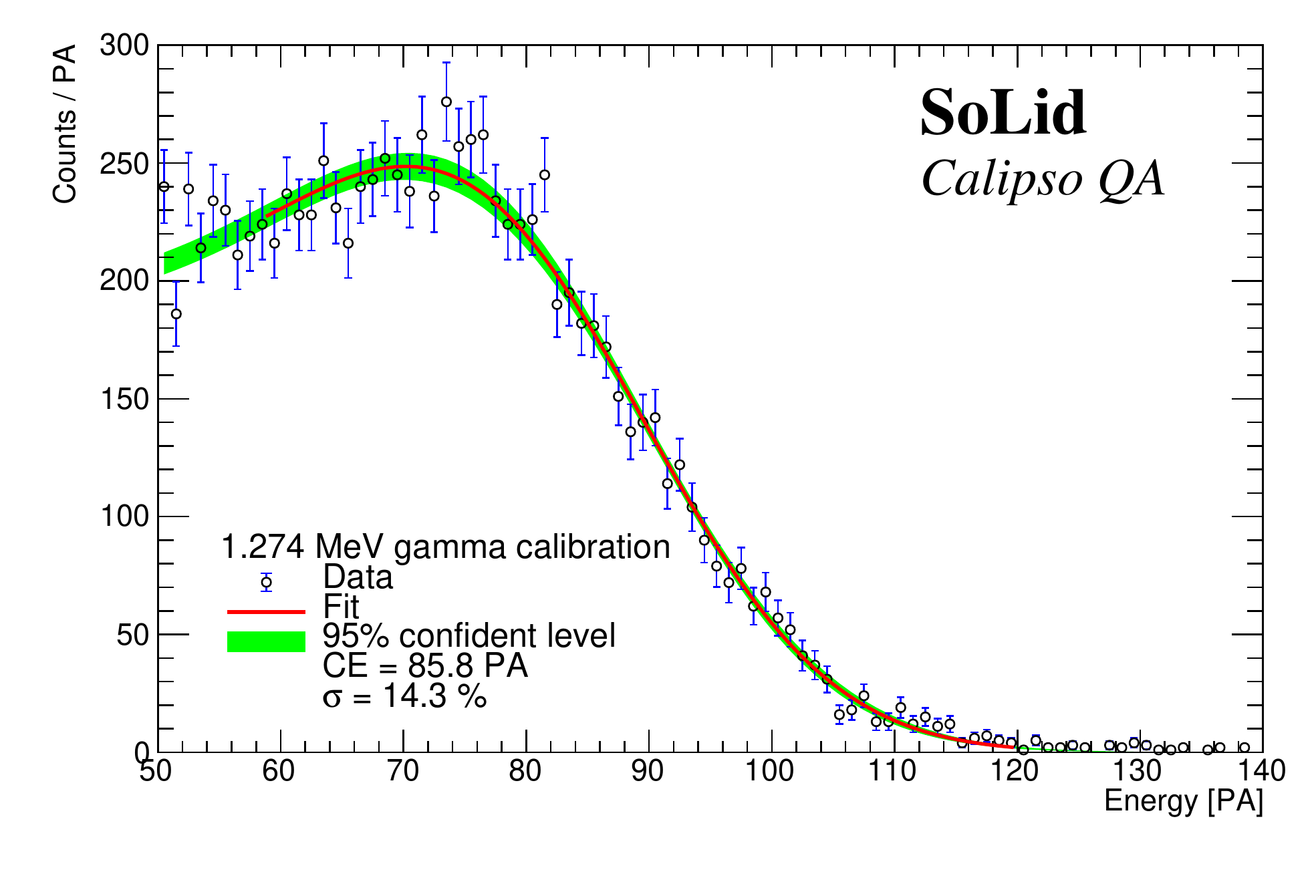}
	\caption{Compton edge profile for a calibration sample in a SoLid cube using a $^{22}$Na gamma source.
	The spectrum is dominated at high energy by the deposits of the 1.27 MeV gamma. The Compton edge is 
	obtained by fitting the distribution with the $pdf$ defined in equation \ref{eqKNpdf}.}
	\label{fig:FitCEDavid}
\end{figure}

Figure \ref{fig:FitCEDavid} shows the energy spectrum for a calibration sample in a SoLid cube. The Compton edge profile
has been fitted using the $pdf$ defined in equation \ref{eqKNpdf}.
The Compton edge is estimated in 85.8 PA respectively, which can be translated in a light yield of 81.2 PA/MeV/Cube. On the other hand, the energy resolution is estimated to be about 14 \%, inline with the SoLid physics requirements.  The accuracy of this fit has been evaluated with the Monte-Carlo. It was found that in the case of the 1.27 MeV gamma from the $^{22}$Na source, the estimation of the CE is biased by about +3.5\%. 
This bias can be explained by cases of multiple scattering in 
the same cube, rather than the assumed single scattering, leaving a deposited energy per cube higher than $E_{CE}$. 

\subsection{Kolmogorov-Smirnov Test}
\begin{figure}[h]
\centering
\begin{tabular}{c }
\includegraphics[width=0.6\linewidth]{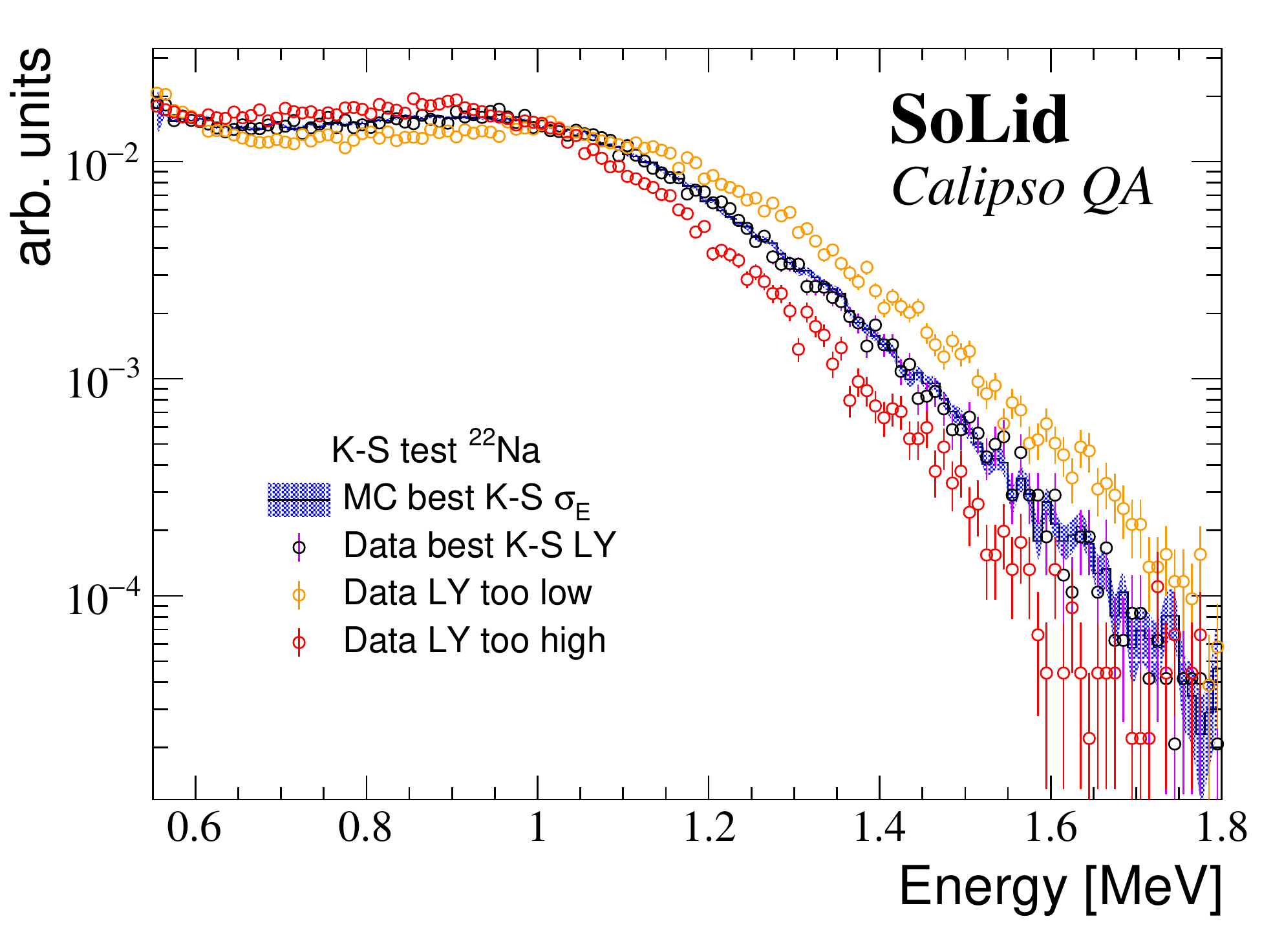} \\
\includegraphics[width=0.6\linewidth]{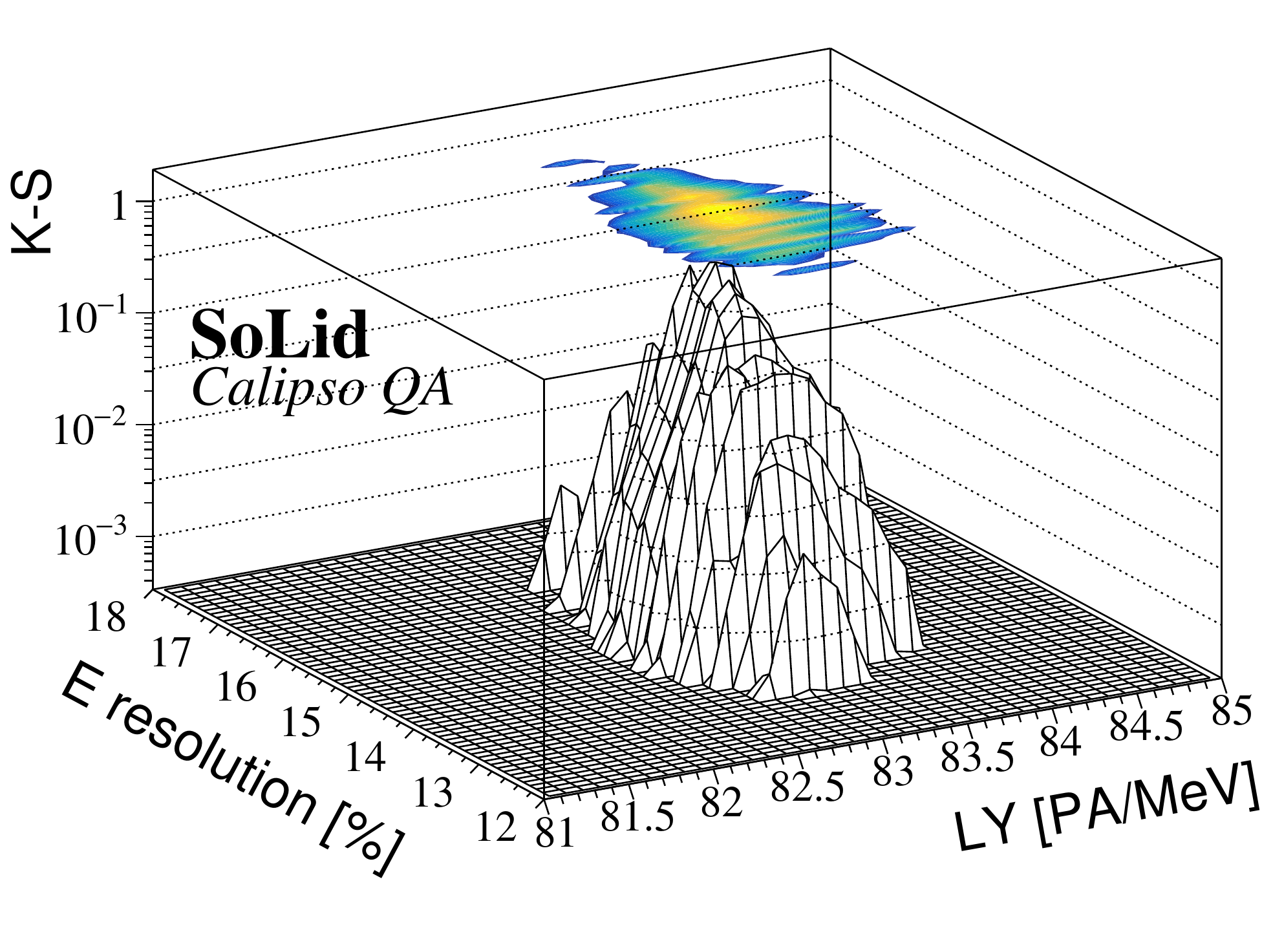}
\end{tabular}
\caption{(Top) Data compared to MC for different values of LY. Violet corresponds to a LY value where the K-S test is
maximised, while the red and yellow show two cases where the LY is found to be too high and too low respectively.
(Bottom) Distribution of the K-S test values in the parameter space of energy resolution from MC and light yield from the data. The K-S test 
takes values of 0 when the data is not compatible with the predicted spectrum, and take positive values when agreement is found.}
\label{fig:KSresult}
\end{figure}

A second light yield estimation is obtained from comparing a {\GEANTfour} simulated
$^{22}$Na energy spectrum with the observed sample using a Kolmogorov-Smirnov (K-S) test.
The true deposited energy in each cube is used to build a set of spectra with different energy resolutions,
varying from 5\% to 20\%. Each spectrum is then compared to the calibration sample for each 
cube, varying the light yield from 50 to 120 PA. The light yield is taken from the point 
of best agreement between the spectra of the calibration sample and the prediction. At this point the K-S test maximises as is shown in Figure \ref{fig:KSresult}.

Figure \ref{fig:KSresult} shows the K-S test results for a specific cube; where the K-S test maximises for a LY of 83 PA/MeV, as shown at the bottom of
figure \ref{fig:KSresult}. 
For values where the data and the Monte-Carlo are not compatible,
the K-S test returns zeroed values.
Varying the section of the spectrum, the binning, and the number of steps used to maximise the K-S test, 
a systematic error of about 2\% was estimated for this method.

Finally the two methods of CE analytical fit and K-S test were compared in order to validate the procedure and provide an estimation of the 
systematic uncertainties. 
Both approaches assume that the convolution product is correct, which means a Gaussian behaviour of the energy resolution at 1 MeV, and that 
the reconstruction efficiency is flat in E. Since only the region around 1 MeV is used, no sizeable effect for introduced errors has been found when using the MC.
All other sources of systematic uncertainties are reasonably assumed to be measured by the difference between the two approaches because they are based on completely different assumptions; the analytical fit supposes that there is only one single 
scattering per cube per event, while the K-S test assumes that the {\GEANTfour} MC is correct.
 
\begin{figure}[!]
  \centering
  \begin{tabular}{c c}
        \includegraphics[width=0.5 \linewidth]{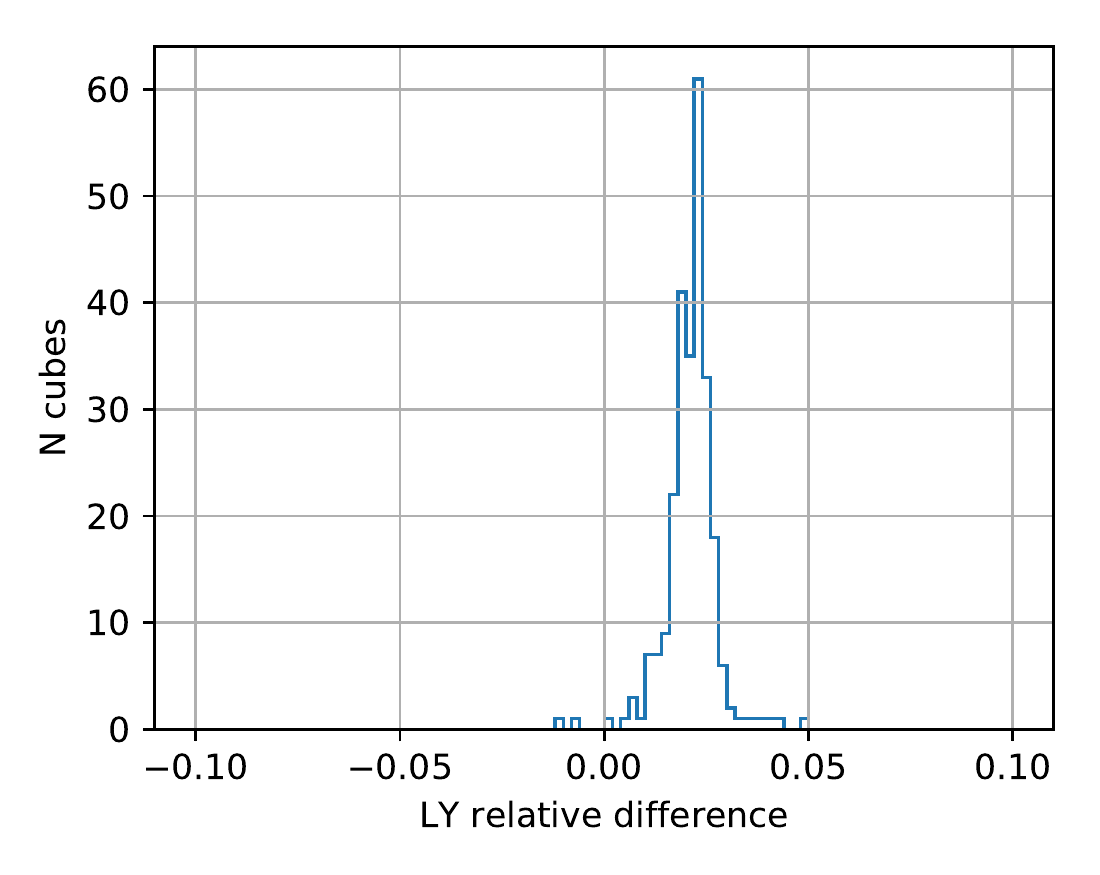}
  \end{tabular}
	\caption{Relative difference in the light yield results for frame 10 using two methods, analytical fit and K-S test.
	A very good agreement can be observed. A difference of about 2\% can be observed, which can be due to the cuts used for the sample selection for the analytical fit.}
	\label{fig:CompareKSFit}
\end{figure}

A very good agreement was found between the values of LY obtained using the K-S test and the  
method of fitting with an analytical Compton $pdf$ after bias correction as shown in figure \ref{fig:CompareKSFit}.
The difference between moth methods remains at less than 2\%. 
The tails on the sides of figure \ref{fig:CompareKSFit} corresponds to cubes with one problematic channel, where the accuracy of the analytical fit is less good. Since the objective of the QA process was only verify  the minimal requirements to achieve the SoLid goals, a 2\% of systematic uncertainty in the energy scale was sufficient at this stage.
The K-S test method was deemed more rigorous when looking to automate the procedure for 
the 12800 cubes, and so this test was used during the QA process prior to the detector construction.

Calibration samples for all 50  frames were collected; as such the LY 
of each SoLid cube was evaluated prior to detector assembly. For the in-situ calibration at BR2, 
a combination of the 2 approaches are continually being used,
providing a good control of systematic uncertainties.

\subsection{Construction Adjustments}
\begin{figure}[!]
	\centering
        \includegraphics[width=0.5 \linewidth]{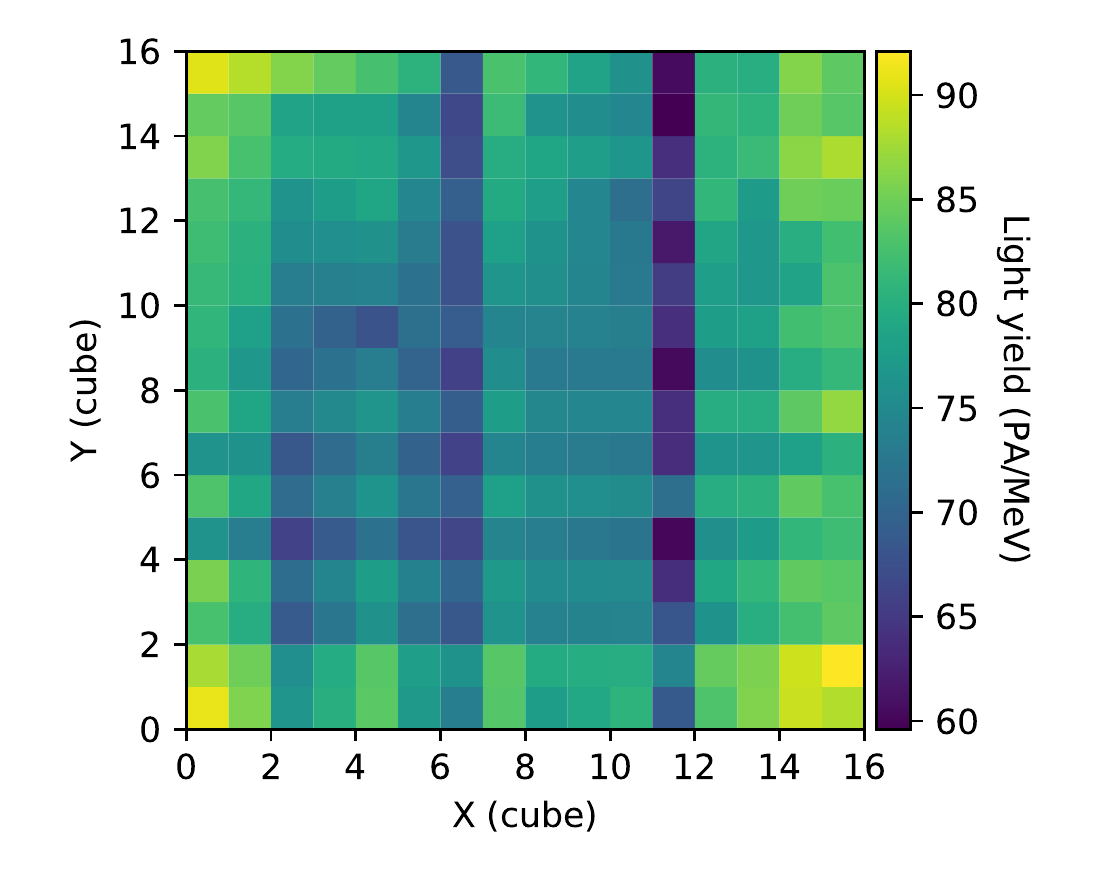}
	\caption{Estimated LY in frame 13 before MPPC crosstalk correction. Column 11 shows a deficit in the LY, which 
	was identified as a bad coupling between the fibre and the MPPC located in the top of the frame.}
	\label{fig:LYFrame13H2}
\end{figure}

Evaluating the light yield for each cube provides a good tool for identification of defective components 
during the construction of the SoLid planes. For example, figure \ref{fig:LYFrame13H2} shows the results of the measured light yield for plane 13,
using the K-S method.  
Looking at this frame, a number of effects can be observed. Firstly, it shows that the cubes placed 
at the border of the frame have a higher LY than those in the centre. This effect is expected if we consider the attenuation of the light within the wavelength shifting fibres used for the readout of the signals, which is estimated to be of the order of $\sim$ 100 cm, which is comparable to the width of one SoLid plane.
Secondly, column 11 shows a light yield that is more than 10\% lower compared to the neighbouring cubes. This low light yield can not be explained in terms of attenuation length, and in cases such as these corrective actions were undertaken. 

In most of these interventions of the row/column showing a low LY, it was observed that the coupling between one of the fibres
and its MPPC or mirror were partially or completely loose. These cases were fixed by either adjusting the connector interface of the MPPC or mirror, or by re-adhering the connector to the aluminium support frame 
used to mount the planes.

\subsection{Light Yield Results}
\begin{figure}[!]
\centering
\begin{tabular}{c c}
  \includegraphics[width=1\linewidth,trim={0cm 0cm 0cm 0cm},clip]{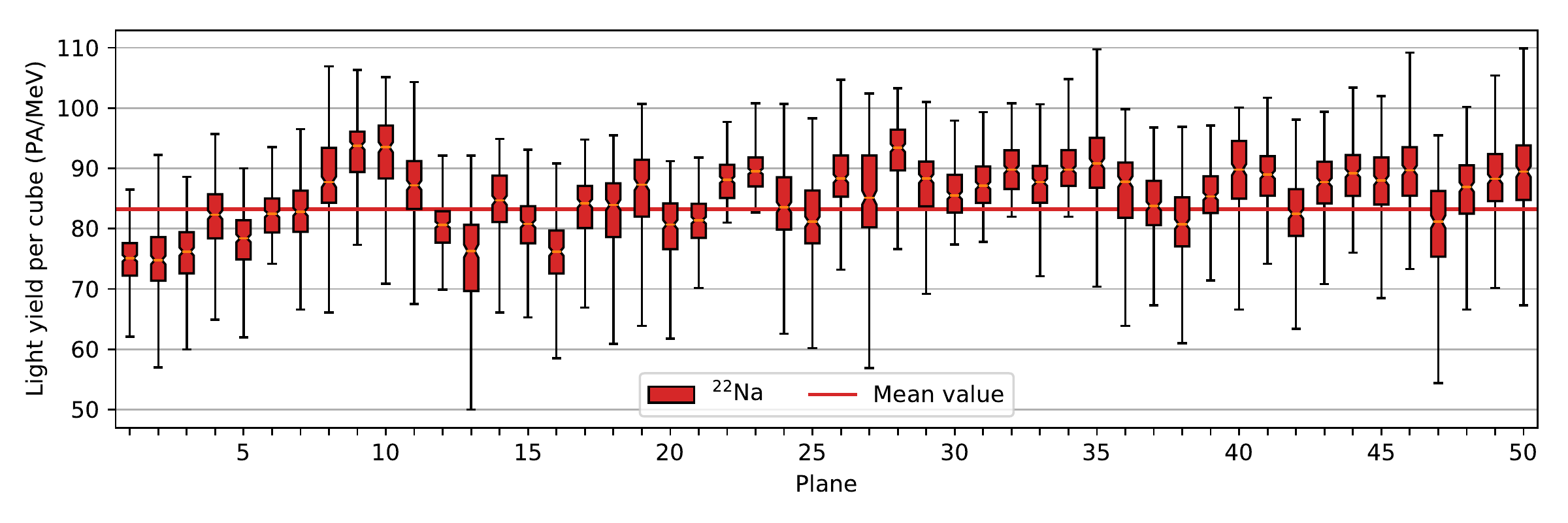} \\
\end{tabular}
\caption{ 
CANDLE plot for the light yield of the 50 planes of the SoLid detector obtained with a $^{22}$Na gamma source. An average of 83 PA/MeV/Cube was found
without MPPC cross-talk subtraction, which is estimated to be around 17\%. Orange line represents the mean value of each plane, while filled boxes represent cubes between the first and the
third quartiles (50\% of the data points). Black lines represent cubes below and above respectively the first
and third quartiles.  }
\label{fig:LY_candle}
\end{figure}

This initial calibration with CALIPSO served not only for quality assurance purposes, but also to obtain a first estimation of the light yield. Consequently a light yield of 83 PA/MeV/Cube before MPPC crosstalk subtraction was measured,  as can be observed in figure \ref{fig:LY_candle}. 
Because of time constraints it was not possible to take dedicated 
crosstalk measurements with the CALIPSO system.
However, the MPPC crosstalk\footnote{This crosstalk 
is  defined as the probability that an avalanching pixel will   cause an avalanche in  a  second pixel.  The  process  happens  instantaneously  an as a consequence, single  incident photons  may occasionally generate signals  equivalent to 2 or 3 photons, or even higher depending on the OV.} has been estimated through other means at $\approx$ 17\% for an OV of 1.5 V \cite{Abreu:2018ajc}.
Therefore the results presented in this paper do not  include correction
for crosstalk. Nevertheless, assuming a MPPC crosstalk of 17\%, the final light yield is expected to be larger than 70 PA/MeV/Cube, inline with the SoLid physics requirements. 

A light yield of about 70 PA/MeV allows to reach an energy resolution of around 12\%. Moreover it could be improved, since the CALIPSO calibration data was taken at an OV of 1.5 V, while the full detector will operate at BR2 at an OV of 1.8 V, increasing the photon detection efficiency by about 20\%.

%% file: sections/Neff.tex
\section{Neutron Detection Efficiency}
\label{sec:Neutron}

The IBD detection efficiency is dominated by the neutron detection
efficiency; therefore the neutron detection parameter needs to be optimised and accurately
determined.

The neutron detection efficiency can be defined as :
\begin{equation}
 \epsilon_{det} = \epsilon_{capt} \times \epsilon_{trig} \times
 \epsilon_{PID}
\end{equation} 
where $\epsilon_{capt}$ is the probability for a neutron to be
captured by the $^{6}$Li in the $^{6}$LiF:ZnS layers, $\epsilon_{trig}$
corresponds to the probability of triggering the read-out on a NS signal
and $\epsilon_{PID}$ is the offline Particle Identification (PID)
efficiency. The last two variables can be factorised as the reconstruction
efficiency $\epsilon_{reco} = \epsilon_{trig} \times \epsilon_{PID}$.
The neutron capture efficiency $\epsilon_{cap}$ is estimated from a Monte-Carlo study
and depends collectively on; the neutron emission point, the energy of the
neutron, the detector geometry, the H content and, the $^{6}$Li content of the $^{6}$LiF:ZnS
layers. Using the CALIPSO based Monte-Carlo configuration, a mean capture efficiency of
$\epsilon_{capt}\approx8.7\%$ has been estimated, with typical variance from 10.6\,\% for a source in the plane centre and
6.5\,\% in the plane corner.


The approach of the QA campaign has been to relatively compare 
each plane, considering that every plane should provide a similar capture
efficiency. Once environmental effects are minimised, a like-for-like comparison can be made on a positional basis. From this the relative efficiency $\epsilon_{rel}$ of each
cube for all given positions is ascertained, while
retaining sensitivity to a large $\epsilon_{capt}$ variation across a plane.
With these positional comparisons; homogeneity, edge effects and other performance related factors can be determined. The absolute efficiency of the entire detector depends on the complete detector geometry and will be determined in-situ after construction is complete at BR2.




\subsection[Nuclear Signal Reconstruction]{Nuclear Signal Reconstruction}

The reconstruction chain from trigger to offline analysis, was also commissioned during the QA process.
In order to reach a high neutron
detection efficiency, the Pulse Shape Discrimination (PSD) is managed in two steps, that are
described in the following sections.
\begin{figure}[!]
\centering
\begin{tabular}{c c}
  \includegraphics[width=0.5\linewidth]{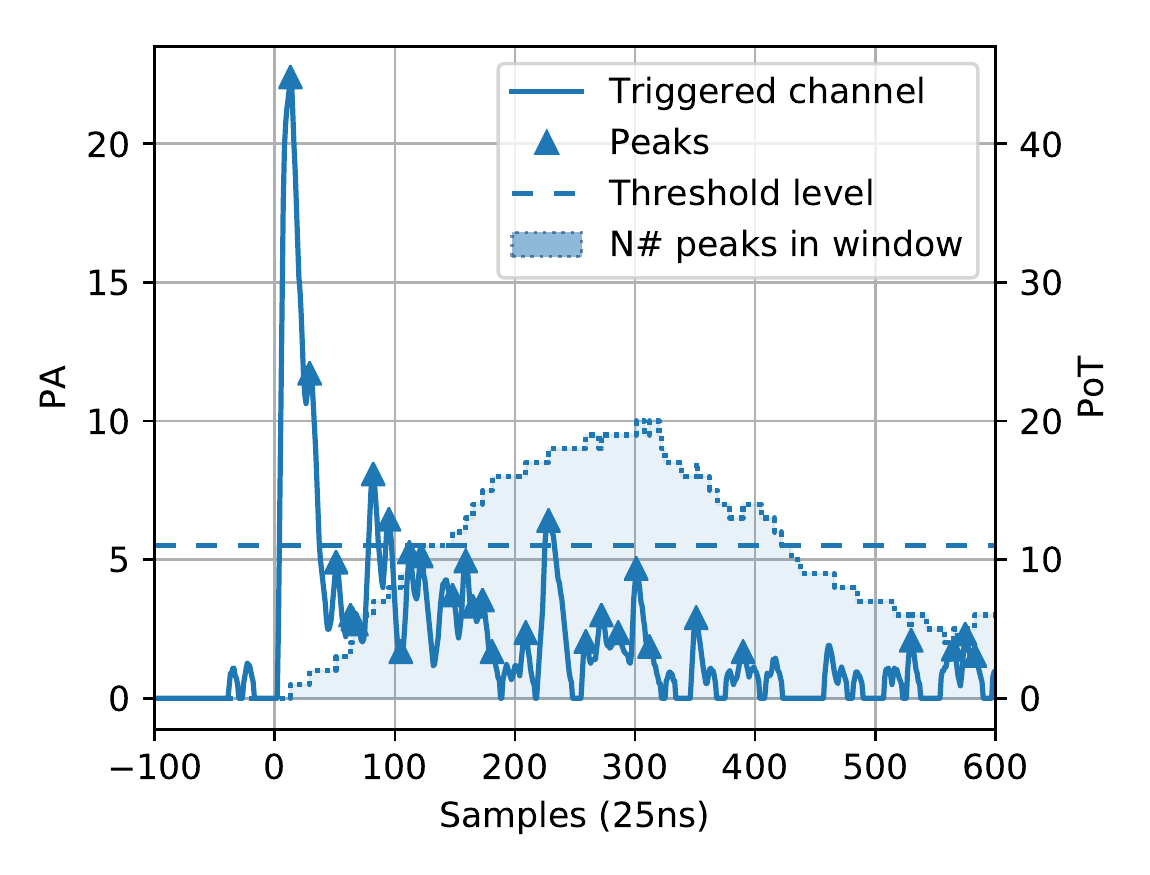} & \includegraphics[width=0.5\linewidth]{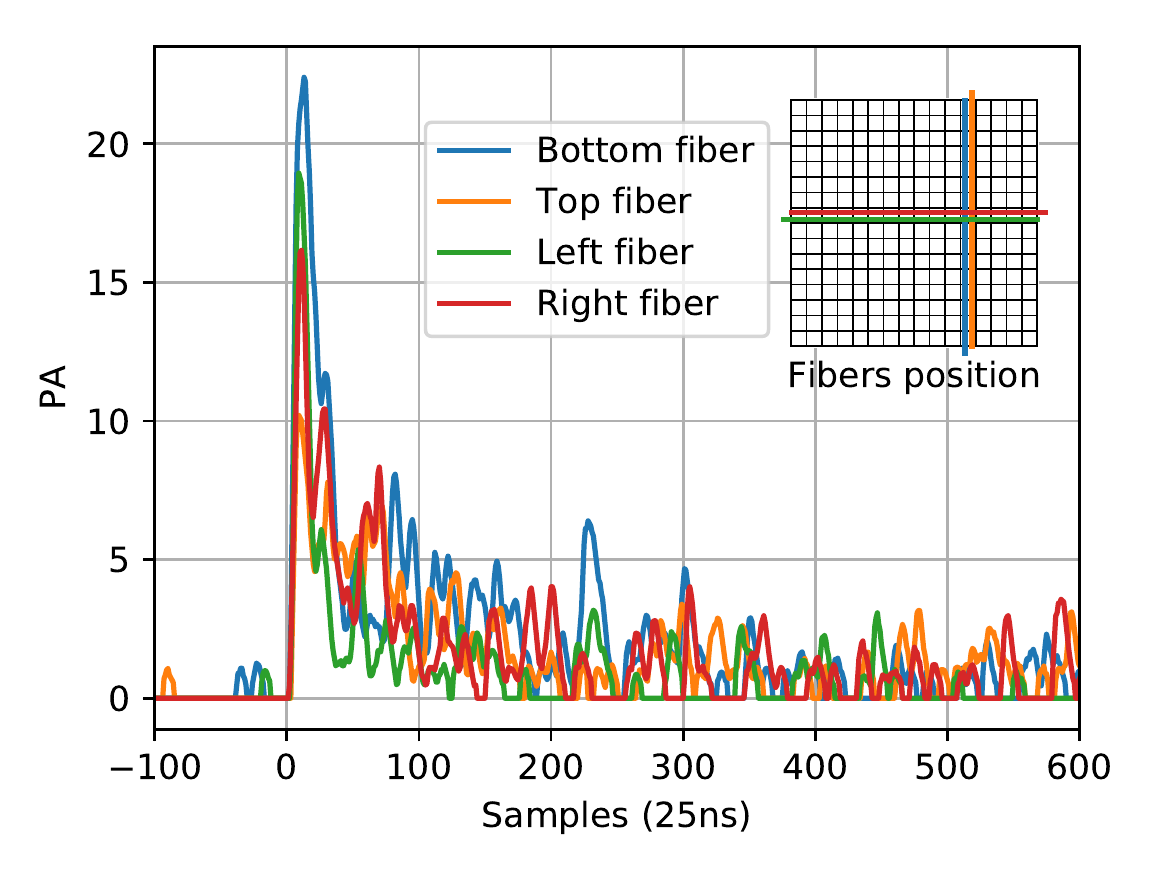}
\end{tabular}
\caption{Left : Triggered NS signal. Peaks (\emph{i.e.} samples above 1.5\,PA and above its two neighbours) are 
represented by triangles. The filled area represents the number of peaks in the previous 256 sample window (Peak over Threshold or PoT). 
Read-out is triggered when the PoT value is above 11 Peaks (dashed line on the figure). \newline Right : Reconstructed NS event, composed from 4 fibres. 
Positions of the fibres in the plane are shown in the upper right insert. From the intersection point, the voxel containing the interaction can be determined.} 
\label{fig:nTrigger}
\end{figure}

\subsubsection[Neutron Trigger]{Neutron Trigger}

The neutron trigger is designed to maximise neutron detection efficiency, even if it is at the expense of the trigger purity. However, 
a high trigger purity can also be achieved by decreasing the neutron detection efficiency.
For instance, increasing the trigger purity from 40\% to 80\% decreases the neutron reconstruction efficiency from 70 \% to 57\% \cite{Abreu:2018njy}. 
The first level of PSD is implemented in the trigger firmware.
This requirement can be reached with a simple
Peaks over Threshold (PoT) algorithm. This PoT algorithm can be implemented since the time constants of PVT and ZnS scintillators
are completely different. An electronic scintillation from the PVT, lasting $\approx$\,20\,ns, is composed of 1 sharp peak; while a nuclear
scintillation from the ZnS, lasting $\approx$\,10\,$\mu$s, is composed of many individual peaks, because of the extended scintillation within the ZnS.  Using this simple PoT approach, the trigger decision is
taken if a certain number of peaks above a given threshold is reached, within
a rolling time window (see figure \ref{fig:nTrigger}).
For the QA campaign, the priority was for a high trigger purity i.e. the number of actual neutrons from the PoT trigger; in this case a  higher threshold was demanded in the trigger requirements (PoT).
In this way a trigger purity exceeding 99\,\% has been achieved. However, this decreased the detection efficiency. For the standard 
data taking in physics mode at BR2, the trigger purity is expected to be $\sim$20\%. This value was initially estimated in order to achieve 
a neutron reconstruction efficiency of around 80\%. Further optimisations increased trigger purity to about 37 \% 
keeping a neutron reconstruction efficiency of about 80\% \cite{Abreu:2018njy}.

\subsubsection[Neutron Particle Identification]{Neutron Particle Identification}

\begin{figure}[!]
\centering
\begin{tabular}{c c}
\includegraphics[width=0.80\linewidth]{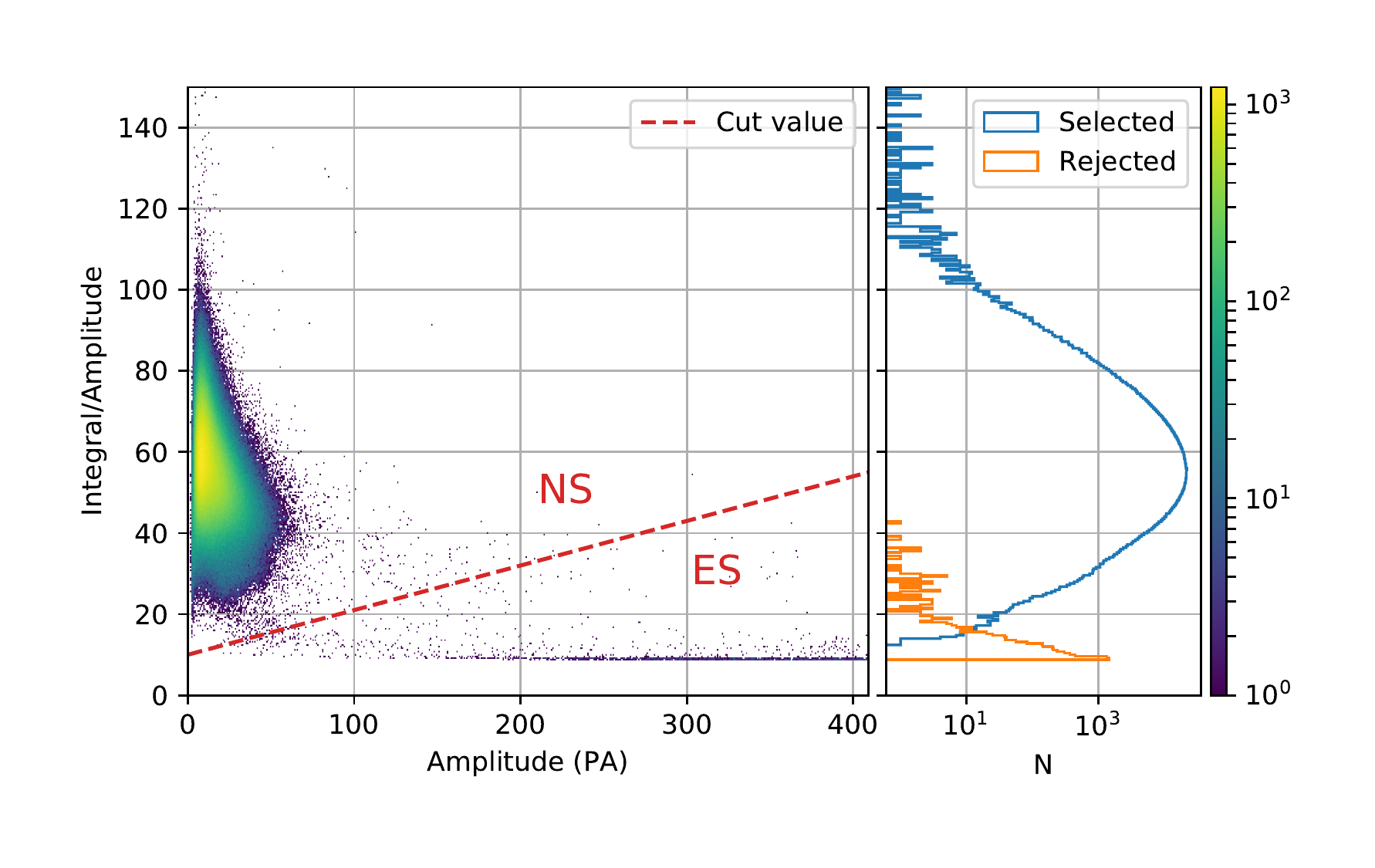}
\end{tabular}
\caption{Integral over Amplitude versus Amplitude, in Photo-Avalanches(PA) for reconstructed
NS candidates from data taken with a $^{252}$Cf source on plane 50. The red dashed line shows the cut used for the Particle Identification. 
The right panel presents the projection on the Integral over Amplitude axis
  for selected and rejected events.}
\label{fig:NS_selection}
\end{figure}

The information from the online neutron trigger is used to initialise the parameters for 
offline analysis of the NS candidates. By finding the combination of 4 channels  that
maximise the number of peaks over threshold in the trigger window, it
is possible to reconstruct the interacting cube position within the plane (see figure \ref{fig:nTrigger}).
Combining the information from these fibres, one can compute the amplitude and integral of the
NS signal. From these two parameters, an NS signal is
placed on the ``Amplitude vs Integral over Amplitude'' (IonA) parameter space, as
shown in figure \ref{fig:NS_selection}. This results in an efficient and
pure separation of ES and NS signals.  It can be noticed that the main
ES contamination in NS candidates is from high amplitude signals, more
likely corresponding to muons crossing the plane.

\subsection[Neutron Detection Efficiency Estimation]{Neutron Detection Efficiency Estimation}

The goal of the relative efficiency
measurement was to compare the number of NS-like events for a given voxel to the
mean number of NS-like events seen throughout all planes, on a positional basis. 

Thus, the number of NS-like events for a given voxel with coordinates x,y in the plane i ($N(x,y,i)$)  can be measured, and the relative efficiency, $\epsilon_{rel}(x,y,i)$, can be defined as:

\begin{equation}
 \epsilon_{rel}(x,y,i) = \frac{N(x,y,i)}{\overline{N}(x,y)}
\end{equation} 
with :
\begin{equation}
  \overline{N}(x,y) = \frac{1}{50}\sum\limits_{i=1}^{50}N(x,y,i)
\end{equation}

Finally, the 25 measurements for each plane are merged in order to
obtain the total number of NS-like events for each of the 256 voxels composing
the plane, as shown in figure \ref{fig:NS_merge}.  However, two
effects need to be taken into account to execute this operation:
\begin{enumerate}
\item{In order to maximally reduce environmental effects,
  only cubes close to the source are added to the global
  measurement.}
\item{The exposure time for each of the 25 calibration positions will have some degree of variance.} 
\end{enumerate}

The first effect is taken into account during the merging process, 
by only selecting cubes within a given distance for X and Y of $\pm$\,3\,cubes, resulting
in a square centred around the source. The total number of
contributing points for each cube can be seen in figure
\ref{fig:NS_merge}.

The number of NS-like events is homogeneous across any given plane, even if only one point of measurement contributes from each of the voxels at the edges of the plane.
Therefore an homogeneous statistical uncertainty over the
whole plane is reached, with minor edge effects on each plane. 
This edge effect can be explained by the
presence of the polyethylene reflector between the cubes and the
aluminium frame. 
Nevertheless, due to the second effect of exposure time variance, it is not
possible to directly compare these merged NS-like maps. Hence, for each
plane, an expected NS-like map is built for this purpose. Thus, for a
given plane $i$, the NS-like map is based on the average rate
observed at each measurement point over all the planes. 
Finally the map is normalised to the time of measurement for the plane $i$.

\begin{figure}[!]
\centering
\begin{tabular}{c c}
  \includegraphics[width=1\linewidth,trim={0cm 0cm 0cm 0cm},clip]{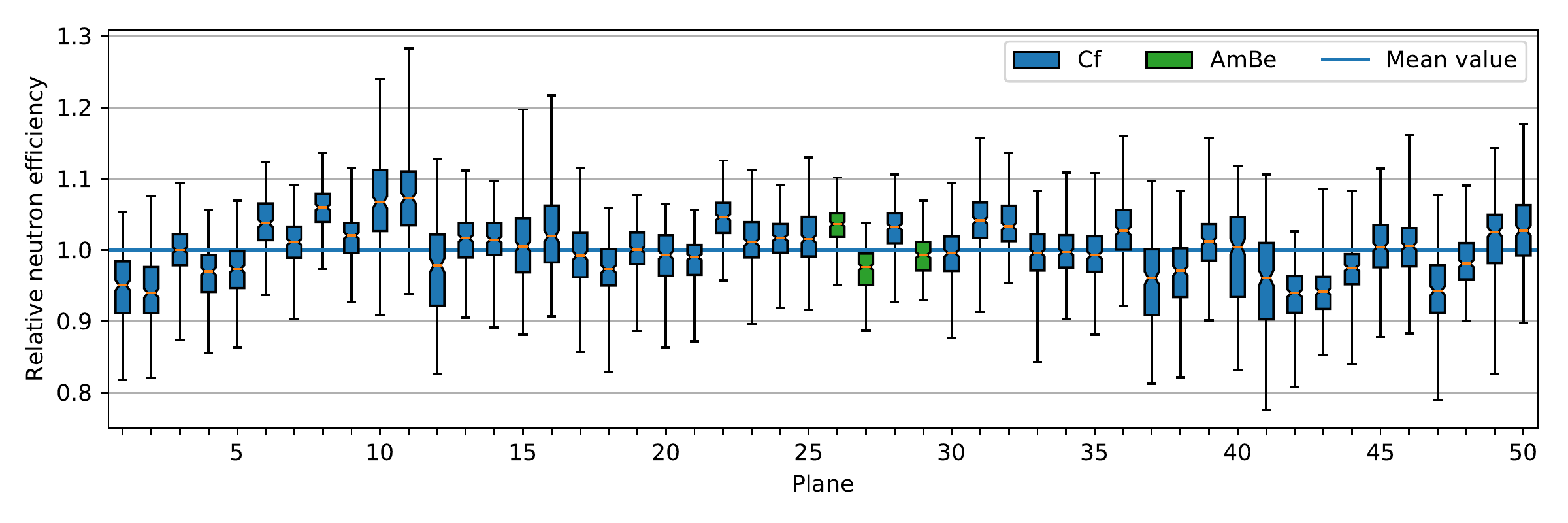} \\
\end{tabular}
\caption{CANDLE plots for the relative neutron efficiency for the 50 planes of the SoLid
  detector. Orange line represents the mean value for each
  plane. Filled boxes represent cubes between the first and the third quartiles (50\% of the data points). Black lines
  represent cubes below and above respectively the first and third quartiles. Results are separated in two
  sets: blue corresponds to planes tested with $^{252}$Cf, green with AmBe.
   }
\label{fig:NS_rel_candle}
\end{figure}

Thus, $\epsilon_{rel}$ for the plane $i$ can be determined as the
ratio of the merged rate for $i$ divided by the mean rate observed in all the planes.
Results for $\epsilon_{rel}$ are presented in figure \ref{fig:NS_rel_candle}. A $\sigma$ of 5\% is found,  
exceeding initial requirements of having a dispersion smaller than 10\%. 

From these measurements, it is possible to extract an absolute
$\epsilon_{reco}$, by comparison to Monte Carlo expectations for the 25
calibrations points.  However, the accuracy in the simulation is not
sufficient to do this exercise per cube, but it can be done to
determine an overall efficiency. This comparison provided an
$\epsilon_{reco} \approx 68.7\,\%$, in-line with the SoLid physics
requirements.  With a neutron purity $\epsilon_{PID}$ above 99\,\%,
which is mainly driven by $\epsilon_{trig}$. Further
optimisations will be done in-situ at BR2, where the trigger will be tuned to
find the best performance metric and achieve a higher neutron detection
efficiency.


\subsection[Construction Adjustments]{Construction Adjustments}
\label{ssec:NrelConsAdj}

The relative neutron detection efficiency measurement identified two possible $^{6}$LiF:ZnS screen related issues, affecting the performance of SoLid.
 
\begin{figure}[!]
\centering
\begin{tabular}{c c}
\includegraphics[width=0.5\linewidth,trim=8mm 0cm 7mm 0cm,clip]{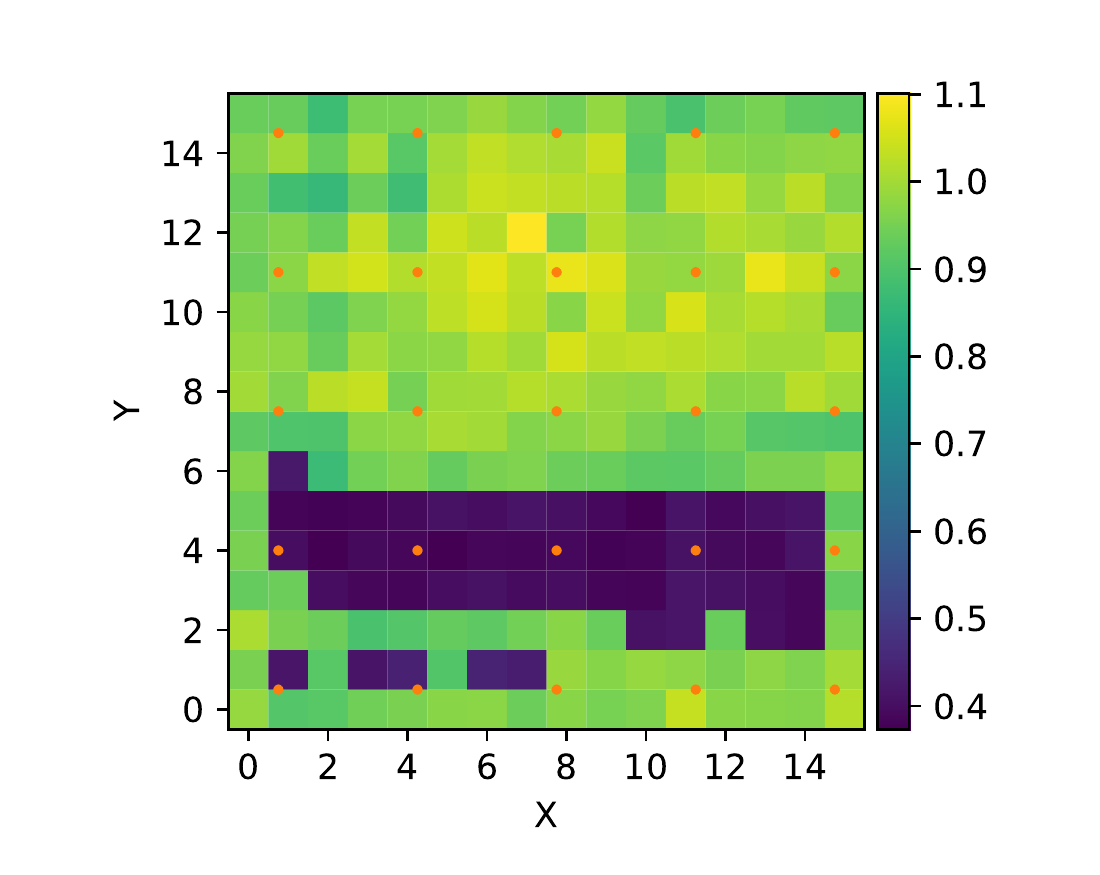}
&
\includegraphics[width=0.5\linewidth,trim=8mm 0cm 7mm 0cm,clip]{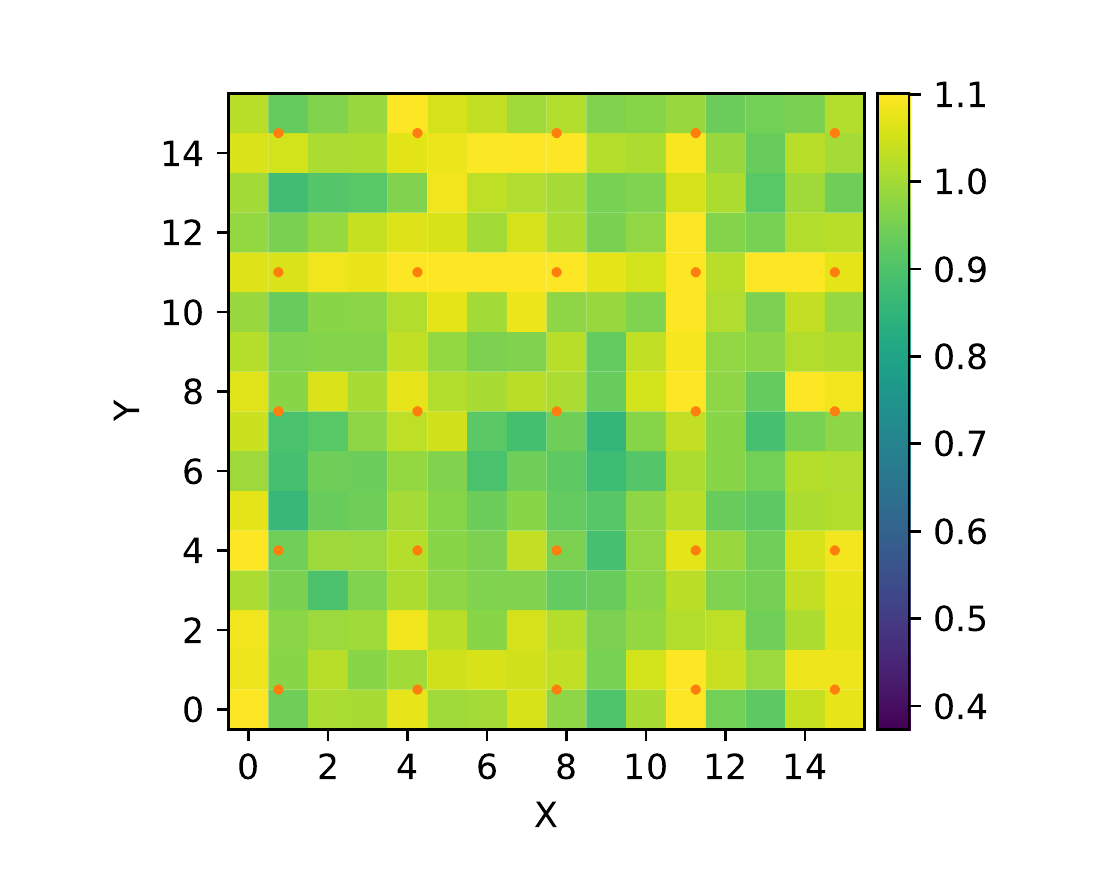}
\end{tabular}
\caption{Orange dots are source positions. Left : Relative NS detection efficiency per cube after data merging for a plane with
  problematic cubes, scanned with Cf. Right : corresponding 
  NS detection efficiency  per cube after replacing cubes containing defective Li screens.
   }
\label{fig:NS_hitmap}
\end{figure}

 Firstly, a particularly low $\epsilon_{rel}$ was observed in some voxels,
whilst the measurements of light yield demonstrated a normal response. See
figure \ref{fig:NS_hitmap}. 
It was determined that these cubes were all
wrapped using $^{6}$LiF:ZnS from a same batch, which was only half doped in $^{6}$Li.  This 50 \% deficit in $^{6}$Li was causing the
lower neutron detection efficiency in these voxels, which were subsequently replaced using new lithium sheets.

The second issue involved one cube from the 12800 voxels
that were tested. It was found that the cube was
wrapped with only one $^{6}$LiF:ZnS screen instead of two, causing
a low neutron detection efficiency.  

The QA process provided an early identification of these two major
issues and a number of minor problems, allowing the time for corrective action prior to assembly at BR2. Thanks to the stringent and all encompassing QA process, an excellent 
performance and homogeneity for the entire detector volume of the SoLid  detector
was achieved.

%
%

Figure \ref{fig:NS_rel_candle} shows the relative NS detection efficiency among the 50 planes of the SoLid detector after replacing cubes containing half doped Li screens. A good homogeneity has been reached in all the planes.

%% file: sections/Conclusions.tex
\section[Discussion and Conclusions]{Discussion and Conclusions}
\label{sec:concl}

A quality assurance procedure was developed and implemented during the construction of the SoLid detector.
For this purpose an automated calibration system called CALIPSO was constructed.
CALIPSO allowed the early identification and fixing of defective components and, provided an initial calibration of the 12800 SoLid detector voxels.
Some minor problems relating to the coupling between the fibres and either the sensors or mirrors were promptly identified and fixed.
In addition, a problem with a batch of $^6$LiF:ZnS screens was identified, which had been half doped with $^{6}$Li.
This problem was rapidly corrected by replacing the problematic
screens with new ones.
Thus, a very good and homogeneous response for all the 12800 SoLid voxels has been achieved. 
This guarantees a consistent operation, uniformity of response and, overall performance of the detector once installed at the BR2 nuclear plant.

This initial calibration for quality assurance purposes provided a first estimation of the light yield and neutron detection 
efficiency in all of the voxels. 
These parameters are expected to be  larger than 60 PA/MeV/cube for light yield and 65\% neutron reconstruction 
efficiency, exceeding the initial requirements 
and previous estimations reported by the SoLid collaboration  \cite{Abreu:2018ajc}. 
Thanks to the QA process, we have shown that the construction of such a novel segmented hybrid detector,
with a total target mass of 1600 kg has been successfully conducted.

The full SoLid detector was commissioned  at the BR2 nuclear plant, at the beginning of 2018. 
The detector is now operational and taking data in stable conditions.